\documentclass[a4paper,fleqn,usenatbib]{mnras}

\usepackage[T1]{fontenc}
\usepackage{ae,aecompl}

\usepackage{graphicx}	
\usepackage{amsmath}	
\usepackage{amssymb}	
\usepackage{subfigure}
\usepackage{multirow}
\usepackage{longtable}

\newcommand{\dinamo}{{\sc dinamo}}
\newcommand{\pcc}{\,{\rm cm}^{-3}}

\newcommand{\um}{\, {\rm \mu m}}

\newcommand{\kel}{\, {\rm K}}

\newcommand{\msun}{\, {\rm M}_\odot}
\newcommand{\nh}{n_{\rm H}}
\newcommand{\nel}{n_{\rm e}}

\newcommand{\pc}{\, {\rm pc}}

\newcommand{\kpc}{\, {\rm kpc}}
\newcommand{\geleven}{G$11.2$-$0.3$}

\newcommand{\gtwentyseven}{G$27.4$+$0.0$}

\newcommand{\gtwentynine}{G$29.7$-$0.3$}

\title[Dust destruction by SNe]{Revisiting the dust destruction efficiency of supernovae}

\author[Priestley et al.]{
F. D. Priestley$^{1}$,
H. Chawner$^{1}$,
M. Matsuura$^{1}$,
I. De Looze$^{2,3}$,
M. J. Barlow$^{3}$,
\newauthor
H. L. Gomez$^{1}$
\\
$^{1}$School of Physics and Astronomy, Cardiff University, Queen's Buildings, The Parade, Cardiff CF24 3AA, UK \\
$^{2}$Sterrenkundig Observatorium, Ghent University, Krijgslaan 281 - S9, 9000 Gent, Belgium\\
$^{3}$Department of Physics and Astronomy, University College London, Gower Street, London WC1E 6BT, UK\\
}

\date{Accepted XXX. Received YYY; in original form ZZZ}

\pubyear{2020}

\begin{document}
\label{firstpage}
\pagerange{\pageref{firstpage}--\pageref{lastpage}}
\maketitle

\begin{abstract}

Dust destruction by supernovae is one of the main processes removing dust from the interstellar medium (ISM). Estimates of the efficiency of this process, both theoretical and observational, typically assume a shock propagating into a homogeneous medium, whereas the ISM possesses significant substructure in reality. We self-consistently model the dust and gas properties of the shocked ISM in three supernova remnants (SNRs), using X-ray and infrared (IR) data combined with corresponding emission models. Collisional heating by gas with properties derived from X-ray observations produces dust temperatures too high to fit the far-IR fluxes from each SNR. An additional colder dust component is required, which has a minimum mass several orders of magnitude larger than that of the warm dust heated by the X-ray emitting gas. Dust-to-gas mass ratios indicate that the majority of the dust in the X-ray emitting material has been destroyed, while the fraction of surviving dust in the cold component is plausibly close to unity. As the cold component makes up virtually all the total dust mass, destruction timescales based on homogeneous models, which cannot account for multiple phases of shocked gas and dust, may be significantly overestimating actual dust destruction efficiencies, and subsequently underestimating grain lifetimes.

\end{abstract}




\section{Introduction}

Since the discovery of large dust masses in some high-redshift galaxies and quasars (e.g. \citealt{bertoldi2003,priddey2003,watson2015}), core-collapse supernovae (CCSNe) have been suggested as a potential source of dust, as they produce the necessary elements for dust formation and occur on short timescales \citep{morgan2003,gall2011}. Far-infrared (IR) observations of nearby supernova remnants (SNRs) in recent years have revealed large ($\gtrsim 0.1 \msun$) masses of newly formed dust in their ejecta \citep{matsuura2011,gomez2012,delooze2017,chawner2019}, lending support to this hypothesis, along with asymmetric line profiles in extragalactic SNe \citep{bevan2016,bevan2019}. However, SNe (of all types) also destroy dust in the ambient interstellar medium (ISM) as the blast wave driven by the explosion sweeps up material, raising the possibility that CCSNe may be net sinks, rather than sources, of dust. Determining the amount of dust destroyed in SN forward shocks is therefore crucial to understanding the evolution of dust in the Universe.

Dust destruction in shockwaves, by both sputtering and grain-grain collisions (shattering), has been studied for decades. However, the large number of physical processes involved, uncertainties in the properties of the dust and gas being modelled, and necessary simplifications such as spherical symmetry, lead to a wide range of reported values for destruction efficiencies, even within the same model. \citet{mckee1987} and \citet{slavin2015} both find order of magnitude differences in the fraction of dust destroyed for relatively modest changes in the magnetic field strength and shock velocity, respectively. In the related topic of dust destruction in SN ejecta by the reverse shock, theoretical destruction rates range from $<1\%$ to $\sim 99\%$ (\citealt{kirchschlager2019} and references therein), and the complication of shock propagation into an evolving ISM, rather than an idealised one-dimensional shock front, has only recently begun to be addressed \citep{hu2019,martinezgonzales2019}.

It is possible to infer the fraction of dust destroyed by measuring changes in the abundances of refractory elements across a shock front, through either absorption \citep{jenkins1976,barlow1977} or emission lines \citep{dopita2016,zhu2019}, or by using IR observations to determine the mass of swept up dust in the shock, which is then in some way converted to a destruction efficiency. This often involves using a model of dust heating in shocks to fit the IR spectral energy distribution (SED), with the amount of dust destruction following from either the required amount of sputtering \citep{arendt2010,sankrit2010,temim2012b}, or the implied dust-to-gas (DTG) mass ratio \citep{borkowski2006,williams2006}, which can be compared to an otherwise known or measured ISM value. IR-based studies typically find destruction efficiencies in the range $20-50\%$, while those based on abundance determinations favour higher values.

The IR-based efficiencies mentioned above are model-dependent, in that in order to calculate the dust emission, the destruction mechanism and shock properties must be assumed, along with the composition and initial size distribution. Observations show that rather than being made up of homogeneous, high-temperature gas as in the shock models, the shocked regions around SNRs additionally contain denser gas which can be identified by molecular emission (e.g. \citealt{reach2005,zhu2014}), which potentially makes up a significant fraction of the total swept up gas mass. \citet{koo2016} found that shock models failed to reproduce the IR to X-ray flux ratios for the majority of SNRs in their sample, suggesting that dust temperatures are lower than would be expected from grains collisionally heated by the shocked gas. \citet{chawner2020} reached similar conclusions, while \citet{zhu2014} required the vast majority of the dust mass in W49B to be in a `warm' component at $\sim 45 \kel$, well below temperatures predicted for collisional heating by shocked gas \citep{dwek1987}. If this situation holds for the majority of SNRs, the dust destruction efficiencies derived from one-dimensional, homogeneous models cannot be relied upon. In this paper, we aim to self-consistently determine the dust and gas properties of a sample of SNRs, assuming the dust is heated by the X-ray emitting gas, and investigate the potential importance of any colder dust component to the inferred destruction efficiency.

\section{Method}

\begin{figure*}
  \centering
  \subfigure{\includegraphics[width=\columnwidth]{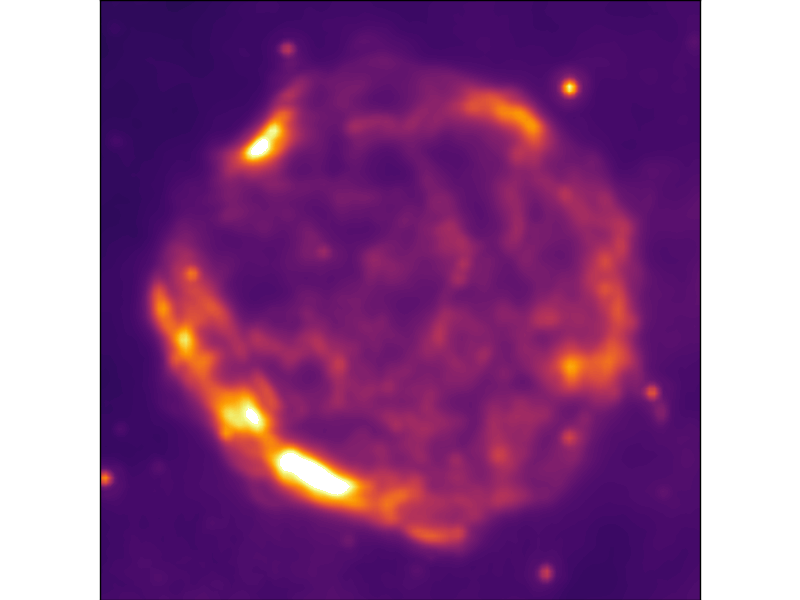}}\quad
  \subfigure{\includegraphics[width=\columnwidth]{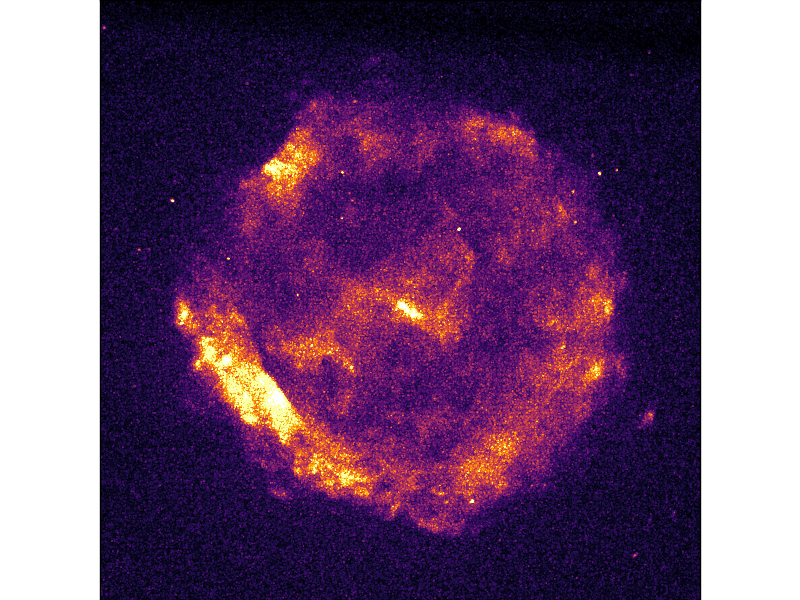}}
  \caption{\geleven{} seen in {\it Spitzer} MIPS $24 \um$ (left) and {\it Chandra} total X-ray flux (right). The image scale is $9.6' \times 9.6'$.}
  \label{fig:g11img}
\end{figure*}

Our method involves first determining the temperature and density of the shocked gas associated with the X-ray emission surrounding an SNR, using these values to calculate dust temperatures from collisional heating, and fitting the IR SED for the same region with the dust emission model. This requires a clear interaction between the SNR and the surrounding ISM in both the X-ray and IR. We thus select objects from the catalogue of SNRs detected by {\it Herschel} \citep{chawner2019,chawner2020} with clear shell-like features in the IR, and for which appropriate archival {\it Chandra} X-ray data is available. The $24 \um$ and {\it Chandra} images of \geleven{} are shown in Figure \ref{fig:g11img}. Emission in both the IR and X-ray is concentrated in a near-circular shell surrounding the central pulsar wind nebula, which is strongly suggestive of relatively uniform interaction with the surrounding ISM, at least compared to other, highly asymmetric SNRs in the \citet{chawner2019,chawner2020} sample. This leaves us with three SNRs - \geleven{}, \gtwentyseven{} and \gtwentynine{}. Their distances and shell parameters (defined below) are listed in Table \ref{tab:snrs}. For the IR data, we use archival {\it Spitzer} and {\it Herschel} data, as well as {\it WISE} bands 3 and 4 to improve coverage in the mid-IR, where shocked dust mostly emits. We describe our modelling process in more detail in the rest of this section.

\begin{table*}
  \centering
  \caption{Names, distances, and adopted shell parameters (defined in text) of the SNRs included in our sample.}
  \begin{tabular}{ccccccc}
    \hline
    Object & $d$/kpc & RA/hh:mm:ss & Dec/$^{\circ}$ & $r_{in}$/${\rm ''}$ & $r_{out}$/$''$ & Ref. \\
    \hline
    \geleven & $4.4$ & $18:11:29.0$ & $-19:25:30.7$ & $88.6$ & $129.2$ & \citet{green2004} \\
    \gtwentyseven & $5.8$ & $18:41:18.4$ & $-4:56:03.5$ & $71.0$ & $102.3$ & \citet{ranasinghe2018} \\
    \gtwentynine & $5.8$ & $18:46:24.1$ & $-2:58:14.5$ & $75.5$ & $114.5$ & \citet{verbiest2012} \\
    \hline
  \end{tabular}
  \label{tab:snrs}
\end{table*}

\subsection{X-ray modelling}

\begin{figure}
  \centering
  \includegraphics[width=\columnwidth]{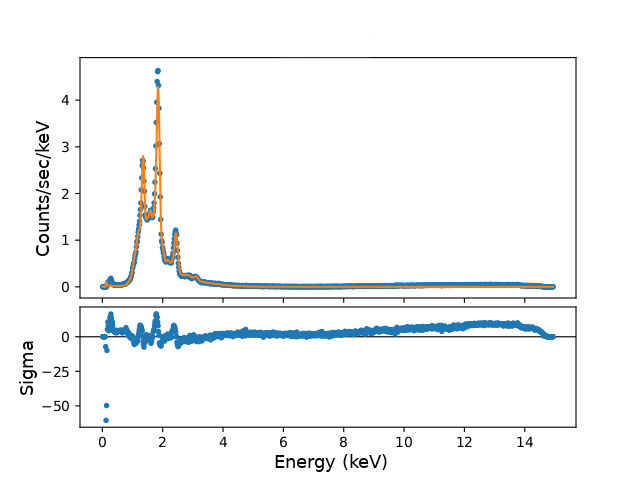}
  \caption{Background-subtracted {\it Chandra} \geleven{} shell spectrum (blue points, top panel), best-fit {\it vpshock*phabs} model (orange line) and residuals (blue points, bottom panel).}
  \label{fig:g11xrayfit}
\end{figure}

For each SNR, we visually inspect the {\it Chandra} image to define a circular annulus containing as much of the shell emission and as little other contamination as possible. The parameters adopted are listed in Table \ref{tab:snrs}. We extract a spectrum from this region using CIAO v$4.12$ (CALDB v$4.9.0$) \citep{fruscione2006}, with a background region selected to be free of any obvious sources. We then fit the spectrum using SHERPA \citep{freeman2001}. Following \citet{borkowski2016}, we use an XSPEC {\it vpshock} model combined with a {\it phabs} photoelectric absorption component. The fit parameters are the temperature $kT$, column density of the absorbing ISM $N_{\rm H}$, and the normalisation constant $C$, while we also allow the abundance of Mg, Si and S to vary, as these elements have prominent lines in all three spectra \citep{borkowski2016}. Figure \ref{fig:g11xrayfit} shows the observations and best-fit model for \geleven{}. The resulting best-fit parameters for each SNR are listed in Table \ref{tab:xrfit}.

\begin{table*}
  \centering
  \caption{Best-fit parameter values using a {\it vpshock*phabs} model to fit the SNR shell X-ray spectra. { Abundances are defined with respect to solar values.}}
  \begin{tabular}{cccccccc}
    \hline
    SNR & $kT$/keV & $N_{\rm H}$/10$^{22}$ cm$^{-2}$ & $C$ & X(Mg) & X(Si) & X(S) & $\chi^2_{\rm red.}$ \\
    \hline
    \geleven & $0.707 \pm 0.004$ & $2.541 \pm 0.006$ & $0.197 \pm 0.003$ & $1.05 \pm 0.01$ & $0.883 \pm 0.007$ & $1.03 \pm 0.01$ & $32.2$ \\
    \gtwentyseven & $0.785 \pm 0.02$ & $2.89 \pm 0.03$ & $0.115 \pm 0.007$ & $1.13 \pm 0.05$ & $0.94 \pm 0.03$ & $1.11 \pm 0.04$ & $2.0$ \\
    \gtwentynine & $2.24 \pm 0.08$ & $2.66 \pm 0.03$ & $0.0107 \pm 0.0004$ & $0.82 \pm 0.05$ & $0.85 \pm 0.03$ & $1.00 \pm 0.04$ & $1.3$ \\
    \hline
  \end{tabular}
  \label{tab:xrfit}
\end{table*}

Our fit for \geleven{} has a much higher $\chi^2_{\rm red.}$ value than \gtwentyseven{} and \gtwentynine{}, despite the relatively small parameter uncertainties. Inspecting the residuals for \geleven{} in Figure \ref{fig:g11xrayfit} shows that this is mostly driven by the emission lines and an unexplained excess at high energies, neither of which should greatly affect the quantities we are interested in. Given that we find almost identical parameter values to \citet{borkowski2016}, we consider the fit reliable. Similarly, our values for \gtwentyseven{} and \gtwentynine{} agree well with the independent measurements of \citet{kumar2014} and \citet{temim2012b} respectively.

Assuming the shell has a constant density, the normalisation parameter $C$ is given by $10^{-14} (4 \pi d^2)^{-1} \nel \nh V$, where $V$ is the volume. If we also assume $\nel \sim \nh$, we can determine the density and thus the gas mass if we know $V$. Assuming the shell is spherical and has a thickness equal to the annular size in Table \ref{tab:snrs}, we can calculate the fraction of the volume which falls within the annulus when projected onto two dimensions, and thus the total volume within the annulus. This gives values of $33.7$, $37.1$ and $59.4 \, {\rm pc^3}$ for \geleven{}, \gtwentyseven{} and \gtwentynine{} respectively. We can then calculate the density from $C$, and, assuming a solar composition, the total gas mass. The derived properties for each SNR are listed in Table \ref{tab:gasprop}.

\begin{table}
  \centering
  \caption{Physical properties of the SNR shells derived from X-ray fits.}
  \begin{tabular}{ccccc}
    \hline
    SNR & $V$/pc$^3$ & $\nh$/cm$^{-3}$ & $T$/$10^6$ K & $M_{\rm gas}$/M$_{\odot}$ \\
    \hline
    \geleven & $33.7$ & $6.8$ & $8.2$ & $7.9$\\
    \gtwentyseven & $37.1$ & $6.5$ & $9.1$ & $8.3$ \\
    \gtwentynine & $59.4$ & $1.6$ & $26.0$ & $3.2$ \\
    \hline
  \end{tabular}
  \label{tab:gasprop}
\end{table}

\subsection{Dust modelling}

With the gas temperature and density known, we can model the dust emission using \dinamo{} \citep{priestley2019}, which calculates the temperature, or temperature distribution in the case of transiently heated grains, for each grain size, accounting for both radiative and collisional heating. We neglect heating by the radiation field, as we do not have any constraints on the flux in the optical and ultraviolet wavelengths important for dust heating, and we expect collisional heating to be the dominant process for these gas properties \citep{priestley2019}.

Whereas previous work has typically assumed an initial grain size distribution, and then evolved it according to a dust evolution model, we use the method from \citet{priestley2020} to determine the size distribution directly. This involves calculating the dust emission for single-size grain models, and fitting the IR SED with the mass of dust at each grain size as the free parameters. This removes any dependence of the result on assumptions about either the dust physics, shock evolution or initial conditions. We choose a three-size model with grain radii $0.001$, $0.01$ and $0.1 \um$, which spans the range of values typically assumed for ISM dust \citep{mathis1977}. Dust in the ISM is generally assumed to be made up of carbon and silicate components, which are affected differently by sputtering and other processes \citep{jones1996}. We do not have sufficient data to simultaneously fit multiple grain sizes for both species, so we consider all-carbon and all-silicate models, which will cover the range of possible results. We assume densities of $1.6$ and $2.5 \, {\rm g \, cm^{-3}}$ for carbon and silicate grains respectively, and use optical properties of BE amorphous carbon \citep{zubko1996} and MgSiO$_3$ silicates \citep{dorschner1995}.

We extract IR fluxes in each band from the same aperture used for the X-ray emission, after masking any point sources and subtracting a median background value taken from the immediate surroundings of the SNR. The issue of determining background flux contributions, or otherwise disentangling emission from SNR-associated and interstellar dust, can significantly affect the results of SED modelling (e.g. \citealt{chawner2019,delooze2019}). We investigate the potential impact of this in Appendix \ref{sec:background}, and note here that it does not substantially change our conclusions. We assume flux calibration uncertainties of $4\%$ for MIPS \citep{engelbracht2007}, $7\%$ for PACS \citep{balog2014}, $5\%$ for SPIRE \citep{bendo2013}, and $4.5\%$ and $5.7\%$ for WISE bands 3 and 4 \citep{jarrett2013}. The fluxes for each SNR are listed in Table \ref{tab:irflux}. We convolve the model SEDs with the appropriate filter response curves to convert to photometric fluxes, and fit the observed fluxes using {\it emcee} \citep{foreman2013}, a Monte Carlo Markov chain (MCMC) code. As this returns a probability distribution function for each parameter (in this case dust mass at a given grain size), we report in Table \ref{tab:dustmass} the best-fit values and the 16th and 84th percentiles as uncertainties.

\begin{table*}
  \centering
  \caption{Background-subtracted IR fluxes extracted from apertures defined in Table \ref{tab:snrs}. All fluxes are in Jy, and band wavelengths are given in $\um$.}
  \begin{tabular}{ccccccccc}
    \hline
    SNR & WISE $12$ & WISE $22$ & MIPS $24$ & PACS $70$ & PACS $160$ & SPIRE $250$ & SPIRE $350$ & SPIRE $500$ \\
    \hline
    \geleven & $4.0 \pm 2.2$ & $30.2 \pm 16.6$ & $26.0 \pm 3.1$ & $124.8 \pm 22.4$ & $176.8 \pm 85.6$ & $65.0 \pm 52.1$ & $33.4 \pm 21.9$ & $15.3 \pm 7.7$ \\
    \gtwentyseven & $0.0 \pm 1.2$ & $8.0 \pm 8.2$ & $13.0 \pm 0.2$ & $25.6 \pm 21.6$ & $124.7 \pm 70.0$ & $94.4 \pm 29.5$ & $47.1 \pm 13.2$ & $16.6 \pm 4.5$ \\
    \gtwentynine & $0.3 \pm 1.6$ & $4.9 \pm 11.2$ & $10.0 \pm 2.1$ & $101.5 \pm 39.1$ & $16.1 \pm 112.4$ & $11.4 \pm 47.4$ & $1.3 \pm 17.3$ & $2.2 \pm 8.2$ \\
    \hline
  \end{tabular}
  \label{tab:irflux}
\end{table*}

\section{Results}

\begin{figure*}
  \centering
  \subfigure{\includegraphics[width=\columnwidth]{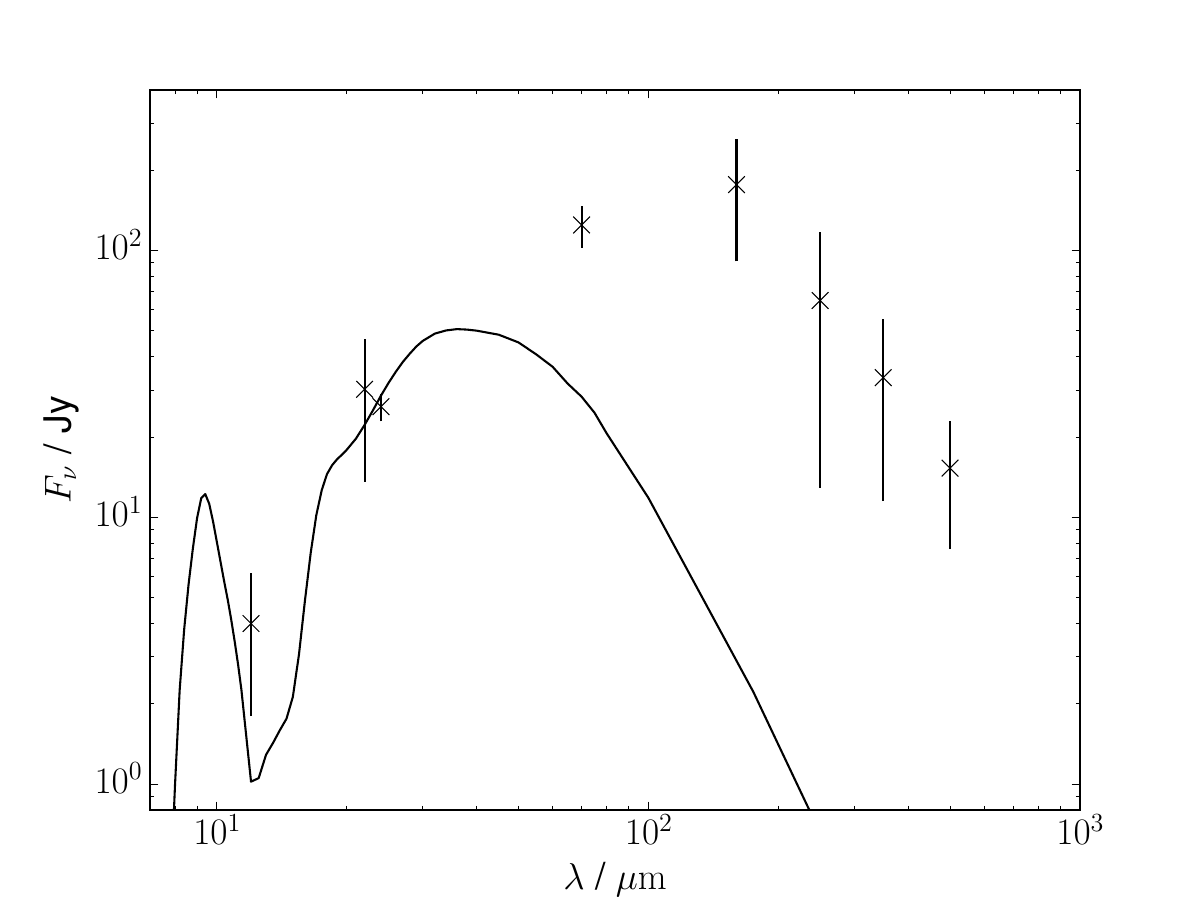}}\quad
  \subfigure{\includegraphics[width=\columnwidth]{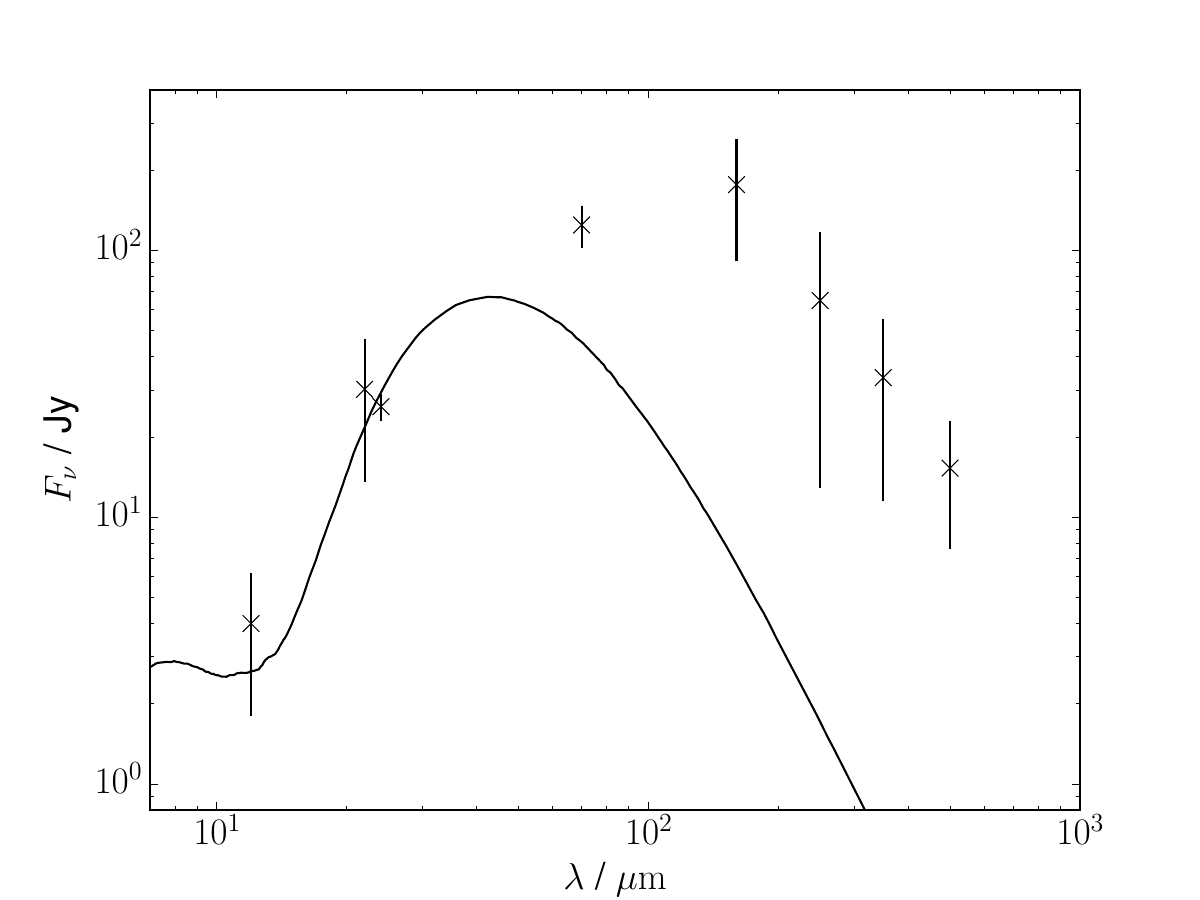}}
  \caption{Best-fit collisionally heated dust SEDs for \geleven{} using silicate (left) and carbon (right) grains (solid line) and observed fluxes (crosses).}
  \label{fig:g11dust}
\end{figure*}

\begin{figure}
  \centering
  \includegraphics[width=\columnwidth]{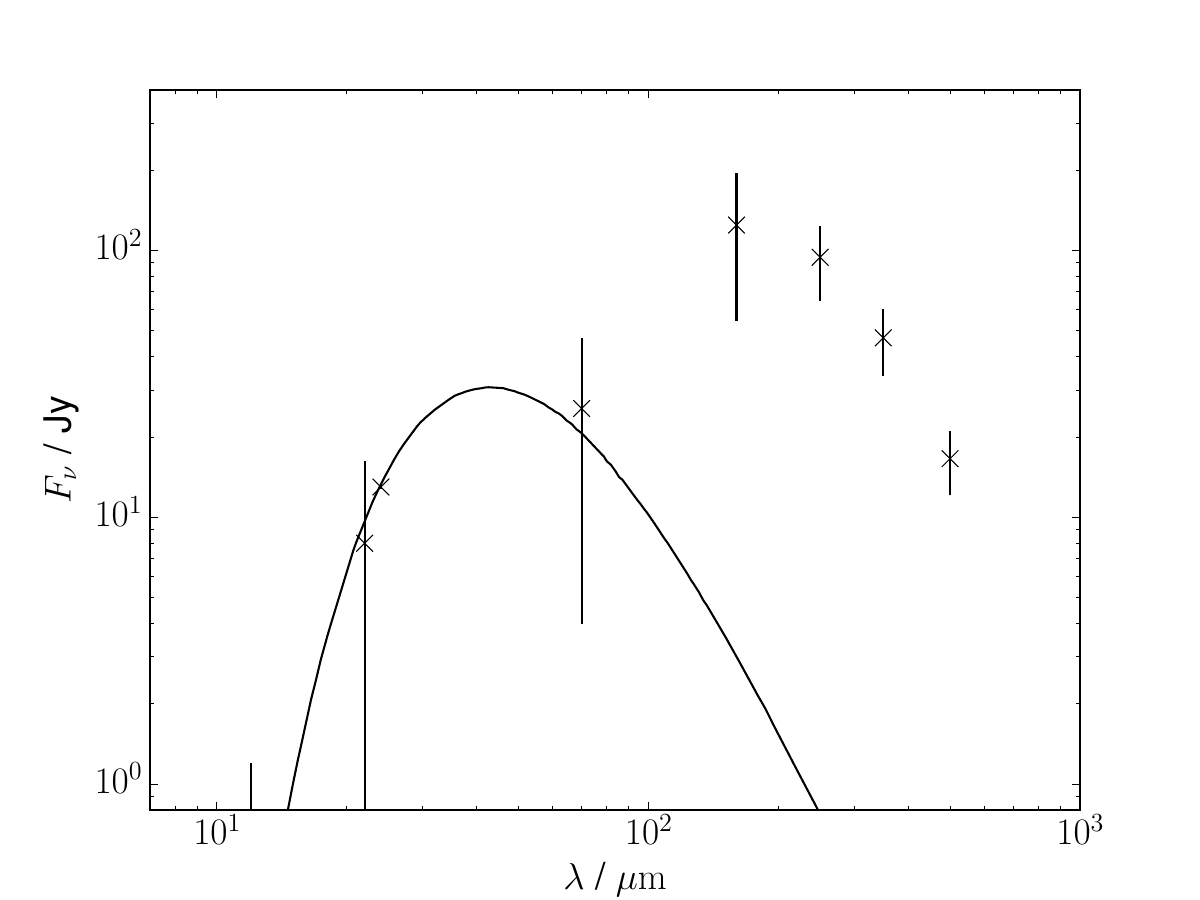}
  \caption{Best-fit collisionally heated dust SED for \gtwentyseven{} using carbon grains (solid line) and observed fluxes (crosses).}
  \label{fig:g27dust}
\end{figure}

\begin{figure}
  \centering
  \includegraphics[width=\columnwidth]{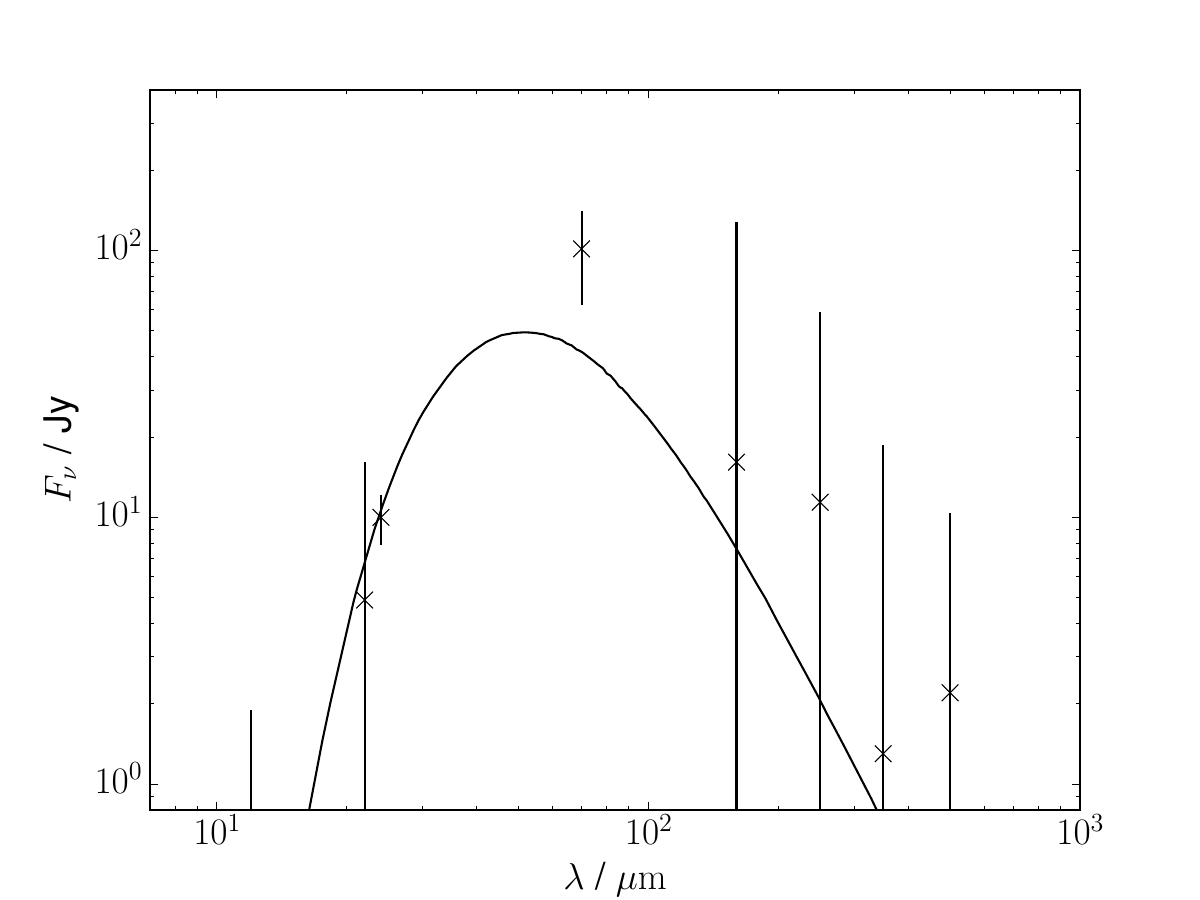}
  \caption{Best-fit collisionally heated dust SED for \gtwentynine{} using carbon grains (solid line) and observed fluxes (crosses).}
  \label{fig:g29dust}
\end{figure}

Figure \ref{fig:g11dust} shows the best-fit collisionally heated dust SEDs for \geleven{}. It is immediately apparent that while the model is a good fit to the mid-IR data, it fails to reproduce the far-IR fluxes for both carbon and silicate grains. Even the $0.1 \um$ grains are heated to high enough temperatures that the $24/70 \um$ flux ratio greatly exceeds the observed value, for either grain composition. Figures \ref{fig:g27dust} and \ref{fig:g29dust} show the carbon grain SEDs for \gtwentyseven{} and \gtwentynine{} respectively. The model can fit the $70 \um$ flux for \gtwentyseven{}, and the far-IR fluxes for \gtwentynine{} are consistent with the data, but both objects clearly require an additional, colder dust component. The clear presence of shell structures in the $70 \um$ images strongly suggests that this is not an issue with background subtraction - the swept up material contains dust which does not appear to be heated by the X-ray emitting gas. Table \ref{tab:dustmass} lists the masses and DTG ratios for each SNR, assuming that all dust is collisionally heated by gas with the paramaters in Table \ref{tab:gasprop}, and the fraction of the total mass in $0.1 \um$ grains. The dust masses and grain size distributions do not vary significantly between carbon and silicate grains for each object. \geleven{} and \gtwentyseven{} both have dust masses of a few $10^{-3} \msun$ and DTG ratios suggesting that only $5-10 \%$ of the initial ISM dust is still present (for an ISM DTG ratio of $0.01$; \citealt{draine2011}). Due to higher dust masses and a lower gas mass, \gtwentynine{} has an inferred surviving fraction of $\sim 50 \%$, similar to the values reported by \citet{williams2006} for SNRs in the Large Magellanic Cloud based on mid-IR observations. In all cases, the fraction of mass in the largest grain size is consistent with unity, suggesting that small grains have been preferentially destroyed. Ignoring the models' failure to fit the far-IR fluxes, these results would suggest highly efficient dust destruction by the X-ray emitting gas.

\begin{table*}
  \centering
  \caption{Dust masses, mass fraction in $0.1 \um$ grains, and DTG ratios for carbon and silicate grains, assuming all dust is collisionally heated by the X-ray emitting plasma.}
  \begin{tabular}{ccccccc}
    \hline
    SNR & $M_{\rm car}$/$10^{-3} \msun$ & $M_{\rm 0.1 \um}$/$M_{\rm tot}$ & $100 \times$DTG & $M_{\rm sil}$/$10^{-3} \msun$ & $M_{\rm 0.1 \um}$/$M_{\rm tot}$ & $100 \times$DTG \\
    \hline
    \geleven & $7.8^{+0.1}_{-1.8}$ & $0.85^{+0.15}_{-0.04}$ & $0.098^{+0.001}_{-0.023}$ & $6.5^{+0.7}_{-1.0}$ & $0.95^{+0.05}_{-0.06}$ & $0.082^{+0.009}_{-0.013}$ \\
    \gtwentyseven & $5.1^{+0.1}_{-2.0}$ & $1.00^{+0.00}_{-0.64}$ & $0.062^{+0.001}_{-0.025}$ & $4.8^{+0.0}_{-1.5}$ & $1.00^{+0.00}_{-0.34}$ & $0.058^{+0.000}_{-0.018}$ \\
    \gtwentynine & $17.8^{+2.6}_{-5.8}$ & $1.00^{+0.00}_{-0.28}$ & $0.556^{+0.082}_{-0.181}$ & $16.0^{+2.0}_{-5.8}$ & $1.00^{+0.00}_{-0.52}$ & $0.501^{+0.063}_{-0.182}$ \\
    \hline
  \end{tabular}
  \label{tab:dustmass}
\end{table*}

Given the failure of models of dust heating by the shocked gas, an additional cold dust component is clearly required. \citet{koo2016} suggested this corresponds to dust located in the denser gas detected via H$_2$ emission in some SNRs \citep{reach2005,zhu2014}, possibly heated by the radiation emitted by the shocked gas. A similar situation exists in Cassiopeia A (Cas A), where the majority of the far-IR flux comes from either unshocked ejecta dust or dense clumps which have passed through the reverse shock, both of which are predominantly heated by the synchrotron emission from the shocked gas \citep{priestley2019}. The SED of the radiation field responsible for the dust heating in the SNRs we consider here is not well constrained, but as the spectral shape is relatively unimportant as long as the total intensity in the optical/UV region is similar, we assume the ISM radiation field SED from \citet{mathis1983} scaled by a factor $G$. We also assume a typical ISM power law size distribution from \citet{mathis1977} (MRN distribution) for this radiatively heated dust to avoid introducing too many additional parameters. Compared to grains located in the high-temperature X-ray emitting gas, the size distribution of dust located in colder gas is presumably relatively unaltered. In any case, the mass opacity of dust in the far-IR is only weakly dependent on grain size, so the resulting dust masses are not greatly affected.

\begin{figure*}
  \centering
  \subfigure{\includegraphics[width=\columnwidth]{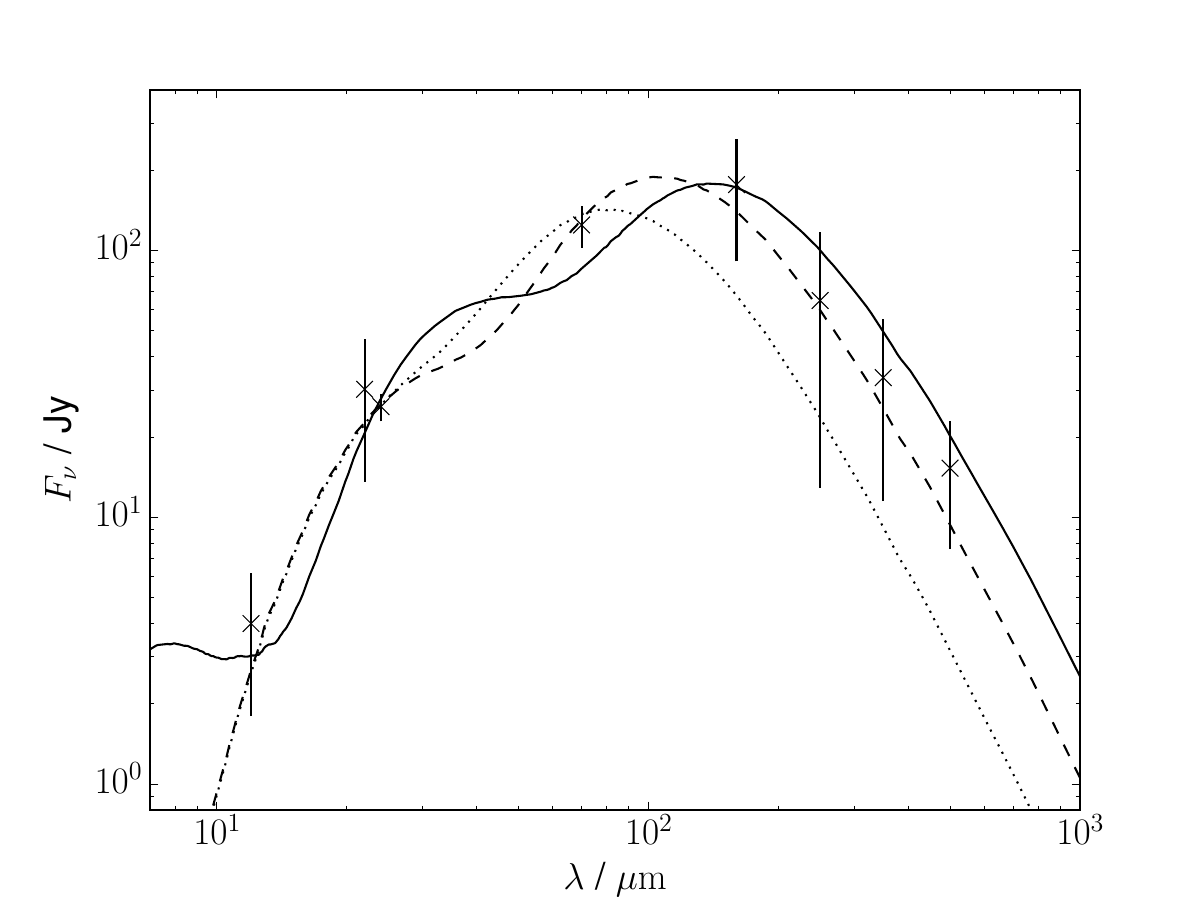}}\quad
  \subfigure{\includegraphics[width=\columnwidth]{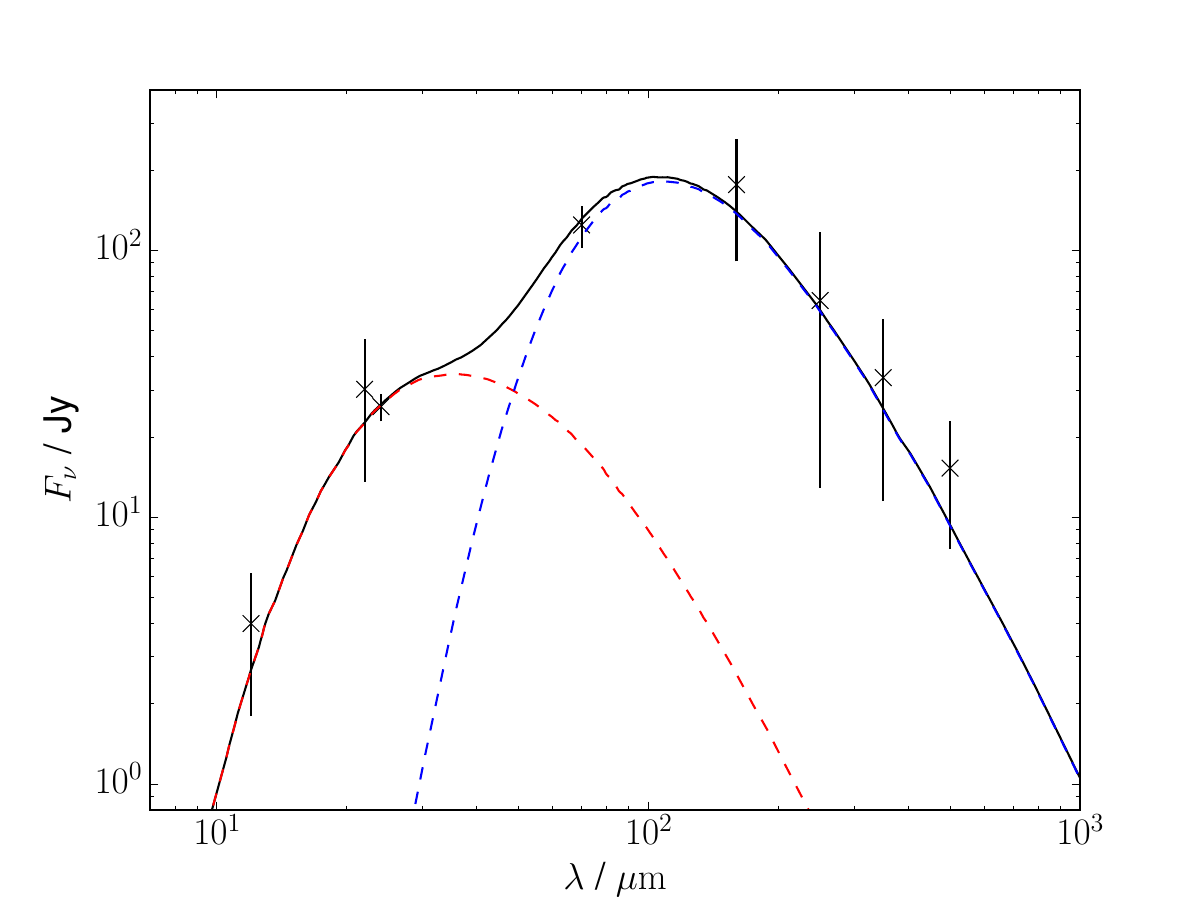}}
  \caption{Left: best-fit carbon dust SEDs for \geleven{}, with both collisionally- and radiatively-heated components, for $G = 1$ (solid line), $5$ (dashed line) and $20$ (dotted line). Right: contribution of the collisionally heated (red dashed line) and radiatively heated (blue dashed line) dust to the total SED (solid black line) for $G=5$.}
  \label{fig:g11cold}
\end{figure*}

\begin{figure}
  \centering
  \includegraphics[width=\columnwidth]{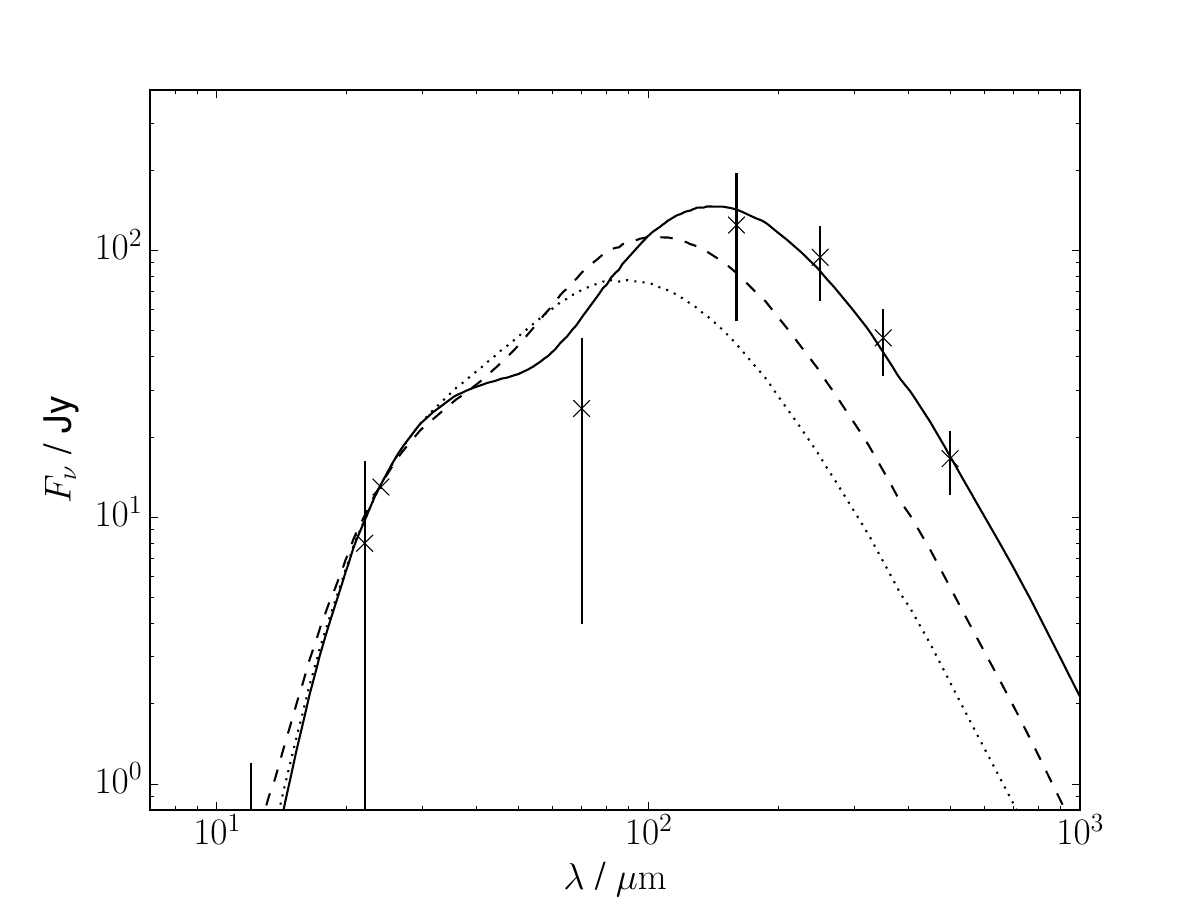}
  \caption{Best-fit carbon dust SEDs for \gtwentyseven{}, with both collisionally- and radiatively-heated components, for $G = 1$ (solid line), $5$ (dashed line) and $10$ (dotted line).}
  \label{fig:g27cold}
\end{figure}

\begin{figure}
  \centering
  \includegraphics[width=\columnwidth]{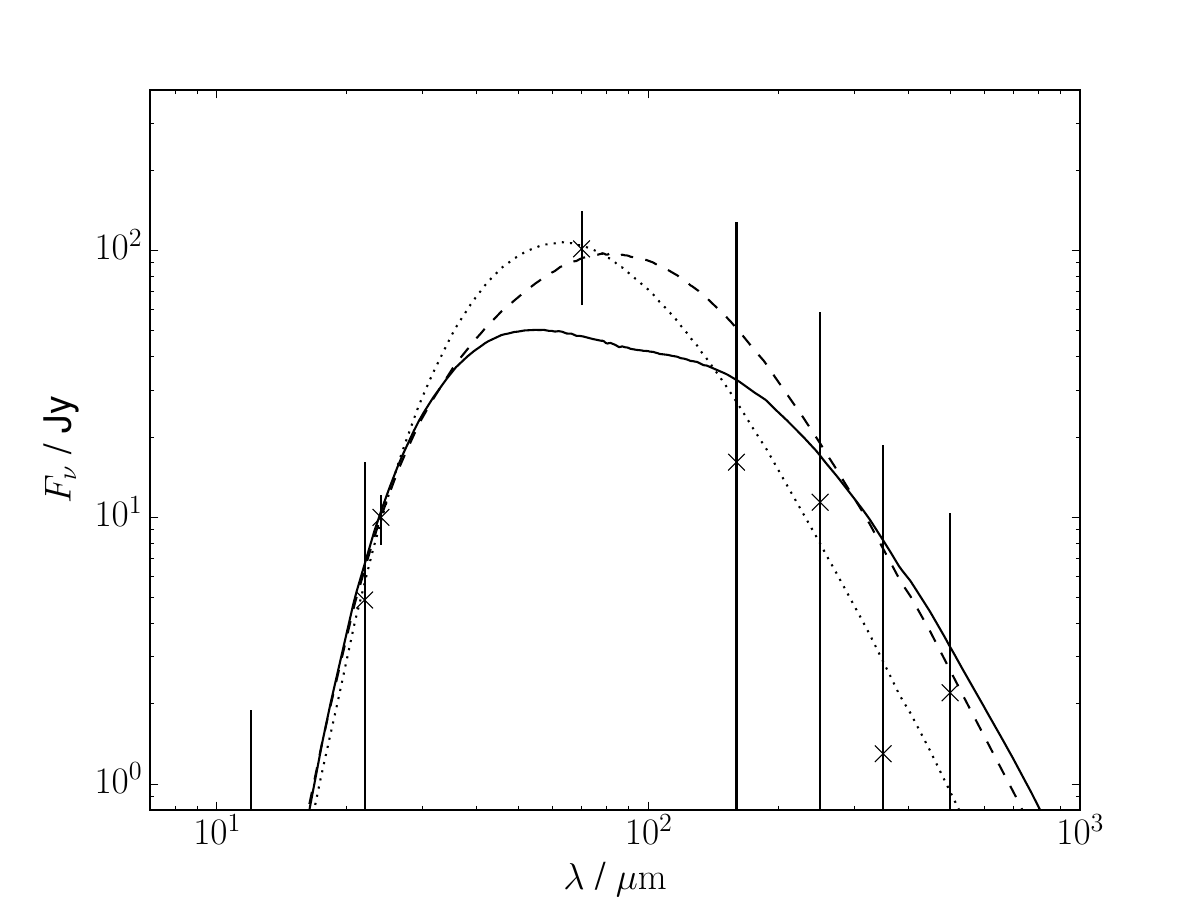}
  \caption{Best-fit carbon dust SEDs for \gtwentynine{}, with both collisionally- and radiatively-heated components, for $G = 1$ (solid line), $10$ (dashed line) and $100$ (dotted line).}
  \label{fig:g29cold}
\end{figure}

We do not attempt to fit $G$ as a free parameter - given the assumptions made in modelling the cold dust emission, it would be difficult to argue that a best-fit value of $G$ has a real physical interpretation. We instead fit the SED of each SNR with warm and cold dust components for values of $G$ in the range $1-100$, and consider the $G$ values giving the lowest $\chi^2$. This allows a reasonable estimate of the required dust masses, for the grain temperatures needed to fit the far-IR SED. Figure \ref{fig:g11cold} shows the resulting fits to the \geleven{} IR SED, for three different values of $G$ with carbon grains, and the relative contributions of the warm and cold dust components to the total SED for $G=5$. We are able to at least roughly constrain $G$ to values $\sim 5$ - for lower values, the addition of colder dust still cannot reach the observed $70 \um$ flux, whereas for higher $G$ fitting the flux beyond $100 \um$ becomes problematic. In any case, all three models require that the vast majority of the dust mass is in the cold component, with masses $0.3-4.6 \msun$ from high to low $G$. The maximium mass of warm dust in any model is a few $10^{-3} \msun$. The situation with silicate grains is similar, although the required $G$ values are higher ($\sim 10$), as silicate grains absorb less efficiently in the optical/UV than carbon.

Unlike \geleven{}, the other two objects do not have a clearly favoured value for $G$. \gtwentyseven{} is best matched by $G=1$, shown in Figure \ref{fig:g27cold}, the lowest value we investigated, and may be even better fit by a lower value. It seems highly implausible that the local radiation field near the forward shock is lower than the typical ISM values, but this is not necessarily what is being implied. The data is better fit by colder dust in the `cold' component, which could equally well be reproduced by a grain size distribution weighted more towards large grains than the MRN distribution we have assumed. As small grains are generally expected to be destroyed more rapidly than large ones \citep{nozawa2007}, this is at least plausible, and arguably more probable. \gtwentynine{} shows the opposite behaviour, with larger values of $G$ providing better fits, up to the maximum value ($100$) that we investigate. The best-fit dust mass in the cold component is still significantly larger than that of the warm dust, but the error bars are consistent with comparable, or even smaller values. However, for values of $G$ up to ten times lower, shown in Figure \ref{fig:g29cold}, the cold dust masses are much larger while the quality of the fit is essentially unchanged. Regardless of what the `true' value of $G$ is, it is clear that the mass of dust in the cold component is plausibly $\gtrsim 0.1 \msun$ and, for \geleven{} and \gtwentyseven{}, likely $\gtrsim 1 \msun$, making the dust mass in this component orders of magnitude larger than the maximum possible in the X-ray emitting gas. The warm and cold dust masses for the $G$ value producing the lowest $\chi^2$ are listed in Table \ref{tab:coldmass} for each SNR. { We also list the DTG ratios in the X-ray emitting gas, assuming only the warm dust is located in this material. We have no corresponding gas mass measurement for the cold dust so are unable to determine DTG ratios, although we estimate probable values in Section \ref{sec:dtg}.}

\begin{table*}
  \centering
  \caption{Dust masses for the best-fitting radiation field strength $G$ { and DTG ratios in the X-ray emitting gas} for each SNR and dust composition, assuming both collsionally heated (`warm') and radiatively heated (`cold') dust components.}
  \begin{tabular}{ccccccc}
    \hline
    SNR & & $G$ & $M_{\rm warm}$/$10^{-3} \msun$ & $100\times$DTG$_{\rm warm}$ & $M_{\rm cold}$/$\msun$ & $\chi^2$ \\
    \hline
    \multirow{2}{*}{\geleven} & carbon & 5 & $2.6^{+3.9}_{-0.3}$ & $0.033^{+0.049}_{-0.004}$ & $1.3^{+0.0}_{-0.3}$ & $1.6$ \\
    & silicate & 10 & $2.8^{+3.5}_{-0.6}$ & $0.035^{+0.044}_{-0.008}$ & $6.7^{+0.3}_{-1.5}$ & $0.6$ \\
    \multirow{2}{*}{\gtwentyseven} & carbon & 1 & $4.9^{+0.2}_{-1.1}$ & $0.059^{+0.002}_{-0.013}$ & $6.7^{+0.7}_{-1.5}$ & $3.5$ \\
    & silicate & 1 & $4.8^{+0.0}_{-1.4}$ & $0.058^{+0.000}_{-0.017}$ & $40.0^{+3.5}_{-9.0}$ & $0.4$ \\
    \multirow{2}{*}{\gtwentynine} & carbon & 100 & $8.3^{+7.3}_{-5.6}$ & $0.259^{+0.228}_{-0.175}$ & $0.090^{+0.034}_{-0.089}$ & $0.1$ \\
    & silicate & 100 & $13.7^{+2.2}_{-4.4}$ & $0.428^{+0.069}_{-0.138}$ & $0.652^{+0.141}_{-0.636}$ & $0.1$ \\
    \hline
  \end{tabular}
  \label{tab:coldmass}
\end{table*}

An alternative explanation for the far-IR excesses is the presence of a cooler X-ray emitting gas component, with correspondingly lower dust temperatures. \citet{temim2012b} suggest this is the case for \gtwentynine{}, where they find evidence for a second X-ray component with $T \sim 10^6 \kel$, an order of magnitude lower than that of the main component ($\sim 1-2 \times 10^7 \kel$ depending on the region analysed). This cooler gas component is also necessary to fit the dust emission. Adding an additional dust emission component from $0.1 \um$ carbon grains, using the same heating parameters as \citet{temim2012b} ($\nel = 2.8 \pcc$, $T = 1.8 \times 10^6 \kel$), we find comparably good fits to the SED as those from our ISM radiation field models (Figure \ref{fig:g29cold}). However, this requires a larger mass of cold dust ($0.13 \msun$) than the $G=100$ model above, so we still find that the warm dust mass is a negligible fraction of the total. The required gas density is also comparable to that of the hot X-ray component, despite the temperature being much lower. \citet{temim2012b} only derive an upper limit on the density, with the value of $2.8 \pcc$ coming from the IR modelling. Higher densities, which would be expected if the two components are in pressure equilibrium and are favoured by several other models from \citet{temim2012b}, produce dust temperatures which are too high to solve the far-IR issue. While it is possible that colder X-ray emitting gas is present, it is difficult to imagine a scenario where it contains the bulk of the dust mass while remaining at densities comparable to, or lower than, the hotter gas.

Another possibility is the presence of larger grains, with typically cooler temperatures which could reproduce the observed ratios of mid- to far-IR flux. We investigate refitting the SED with the addition of $1 \um$ grains, which were found to be necessary for similar reasons in \citet{priestley2020}. For \geleven{} and \gtwentyseven{}, these models still fail to fit the data beyond $100 \um$, although for \geleven{} the $70 \um$ flux can now be reproduced. For \gtwentynine{}, a good fit to the entire SED is possible, if unsurprising given the large observational uncertainties. However, for all three SNRs this requires that the majority of the dust mass ($\gtrsim 90 \%$) is made up of micron-sized grains, and the DTG ratios are greater than our assumed ISM value of $0.01$ for \geleven{} and \gtwentynine{}. Similar conclusions regarding large grains were reached by \citet{priestley2020}, but that related to SNR ejecta dust, for which large grain sizes have previously been found necessary (e.g. \citealt{wesson2015,bevan2016}). For preexisting ISM dust, we are unaware of any proposed model in which the size distribution is this top-heavy and extends to such large grain sizes - extending the MRN power law distribution to $1 \um$, the fraction in micron-sized grains is $\sim 30 \%$. Models of ISM dust growth \citep{kohler2015,ysard2016} do predict the existence of larger grains, but the total dust mass is still mostly made up $\sim 0.1 \um$ radius grains. While preferential destruction of smaller grains by sputtering \citep{nozawa2007} would be expected to increase this value, the original dust mass would then have to be higher by a factor of $\sim 3$, making the implied initial DTG ratios higher than that of the typical ISM for \gtwentyseven{}, and larger than the total mass of swept-up metals for \geleven{} and \gtwentynine{} (assuming solar metallicity). Given that the addition of micron-sized grains only manages to fit two additional data points across the three SNRs, we do not consider it a plausible explanation for the far-IR flux excesses.

\section{Molecular emission}
\label{sec:mol}

If the excess far-IR emission for these three SNRs originates from dust in high-density shocked gas with much lower temperatures, we would expect to see evidence for this in molecular line emission. In fact, \citet{kilpatrick2016} find that all three SNRs show broad, high-velocity CO $J=2-1$ lines, which they attribute to interaction between the SNR and surrounding molecular gas. While reassuring, CO emission can also be produced by unassociated clouds along the line of sight, and the presence of high velocity CO emission does not necessarily mean that gas with the required properties exists in large quantities. Near-IR H$_2$ emission, on the other hand, must be produced by dense, warm gas (e.g. \citealt{priestley2018}), as colder material cannot excite the upper levels, while at X-ray emitting temperatures there are insufficient quantities of surviving H$_2$. From our sample, \geleven{} has been detected in multiple H$_2$ transitions \citep{koo2007,andersen2011}, allowing us to investigate the physical conditions in this molecular-emitting gas.

\begin{figure}
  \centering
  \includegraphics[width=\columnwidth]{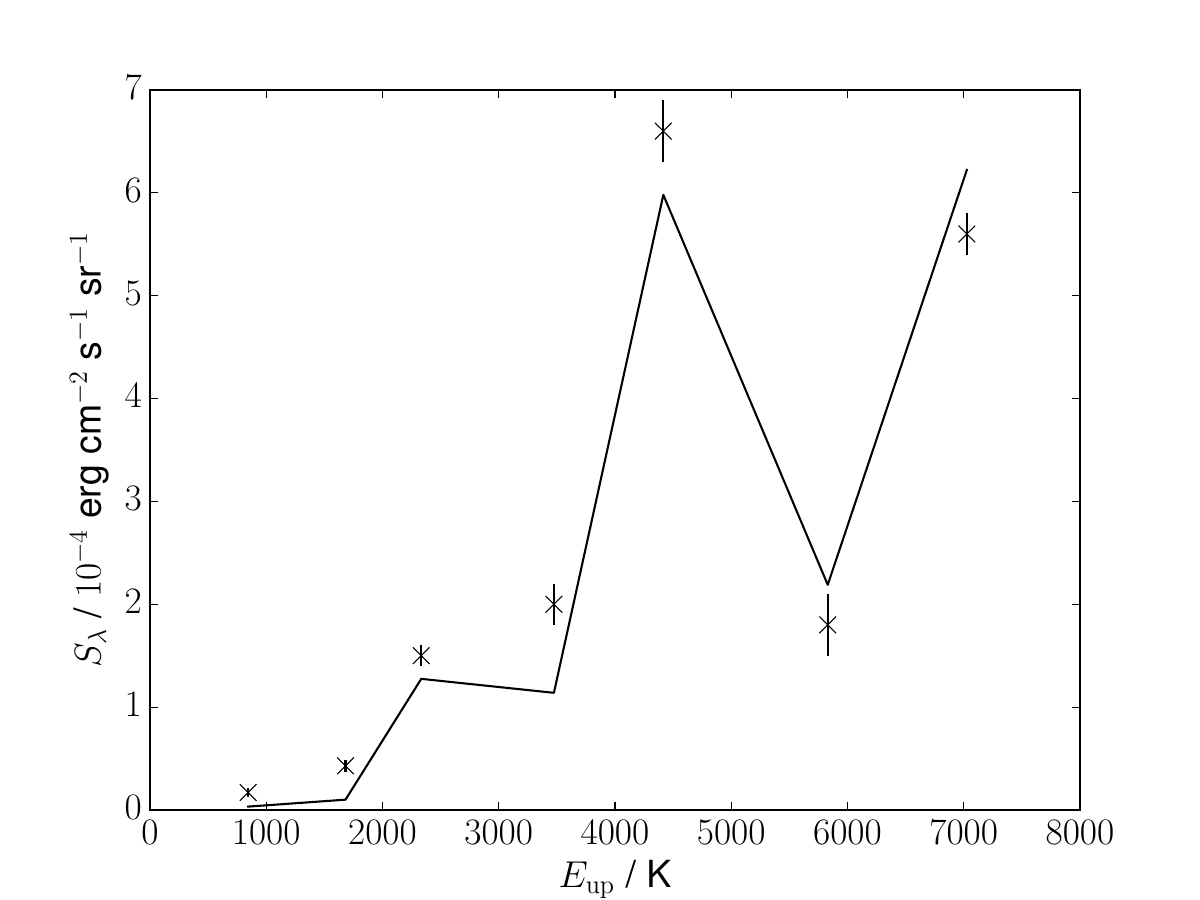}
  \caption{Observed H$_2$ line fluxes versus upper level energy for \geleven{} (black crosses), and values for a $5000 \kel$, $10^3 \pcc$ model with a depth of $0.84 \pc$ (solid line).}
  \label{fig:h2emis}
\end{figure}

We use {\sc ucl\_pdr} \citep{priestley2017}, a one-dimensional photodissociation region (PDR) code, to calculate the H$_2$ line emissivities for densities and temperatures in the range $10-10^5 \pcc$ and $1000-10^4 \kel$. The code calculates the gas-phase chemistry and level populations self-consistently, so the only free parameter to obtain line fluxes is the `depth' of emitting material, as the relative line strengths are constant for fixed gas properties. Taking the H$_2$ S(1)-S(7) line surface brightnesses for \geleven{} from \citet{andersen2011}, we find the ratios are best reproduced by $n = 10^3 \pcc$ and $T = 5000 \kel$, shown in Figure \ref{fig:h2emis}. Moreover, the `depth' of this material required to match the observed fluxes is $0.84 \pc$, compared to our assumed shell thickness of $0.86 \pc$, in excellent agreement. While we have not exhaustively searched the available parameter space, it is clear that, for this SNR at least, the presence of the required quantity of warm, dense gas is entirely consistent with the molecular observations, and in fact it would be strange if the H$_2$ emitting gas did not contain dust at colder temperatures than the grains heated by the X-ray emitting plasma.

\section{Discussion}

\subsection{Dust destruction efficiencies}
\label{sec:dtg}

Our results are consistent with those of \citet{koo2016}, in that the observed IR emission requires colder dust temperatures than predicted by shock models, and \citet{zhu2014}, as the cold dust component dominates the overall mass budget. This work represents an improvement over the previous results, as we make use of the full far-IR SED up to $500 \um$, and directly calculate the grain temperatures from the measured gas properties rather than simply fitting the SED with a blackbody. We also account for different grain sizes reaching different temperatures, including the stochastic heating of small grains. Whatever the nature of the cold dust component, it is clear that collisionally heated dust grains located in the X-ray emitting gas are incapable of reproducing the full IR SED.

The mass of cold dust required depends on the grain composition, size distribution and heating mechanism, all of which are unknown and difficult to constrain with limited far-IR photometry. While we have assumed the grains are heated by radiation from the shocked gas, for the gas properties derived in Section \ref{sec:mol}, collisional heating may have a non-negligible contribution to the heating rate. However, for the grain temperatures required to fit the far-IR photometry, cold dust masses are of order $1 \msun$ for \geleven{} and \gtwentyseven{}, and potentially slightly lower for \gtwentynine{}, for any reasonable dust opacity. For an angular radius $100''$ and distance $5 \kpc$, the mass of gas swept up by an SNR is $\sim 50 \left(n/100 \pcc\right) \msun$, where $n$ is the average number density of the pre-SN gas. For an ISM DTG ratio of $0.01$, the inferred dust masses require an associated gas mass corresponding to initial densities of a few hundred $\pcc$, consistent with those of molecular clouds. As the three SNRs we study were core-collapse explosions from massive stars, it is reasonable that they are still located near to their birth clouds. \citet{reach2005} inferred higher gas densities ($\gtrsim 10^3 \pcc$) in the molecular component for the objects they investigated, which can be reconciled with our estimate with a filling factor below unity. This is required anyway, as the majority of the shocked volume seems to be made up of the lower density gas.

The DTG ratios for the X-ray emitting gas and the associated warm dust, listed in Table \ref{tab:dustmass}, suggest very high destruction efficiencies ($\gtrsim 90\%$), except for in \gtwentynine{} where approximately half of the original dust has been destroyed. However, the mass of both dust and gas in this phase is negligible compared to the total swept up mass, so the true overall destruction efficiency will be effectively equal to that of the cold component. If the entire volume into which \geleven{} has expanded was initially filled with $10^3 \pcc$ gas (the best-fit density for the H$_2$ emission), the gas mass would be $\sim 800 \msun$, which suggests surviving dust fractions of $\sim 20 \% \, (80 \%)$ for the carbon (silicate) masses in Table \ref{tab:coldmass}. As the filling factor of the dense gas is certainly well below $1$, these values represent lower limits to the true surviving fraction. \citet{priestley2019} measured DTG ratios in pre- and post-shock ejecta for Cas A and found approximately half the dust in dense clumps had survived the passage of the reverse shock, compared to almost complete destruction in the diffuse material. Assuming the underlying physics does not differ significantly, this is a reasonable guess as to the true ISM value. Further molecular line observations towards the other SNRs would be useful for better constraining this value.

\subsection{Comparison to previous work}

Assuming a destruction efficiency of $\sim 50\%$ is appropriate for the cold component, our results are in reasonable agreement with previous determinations of this value from model fits to IR data \citep{williams2006,borkowski2006,sankrit2010,arendt2010}, despite these studies assuming the dust emission comes from the X-ray emitting gas. There are two factors explaining this: firstly, if the vast majority of the gas mass is also cold ($\lesssim 10^4 \kel$; this presumably must be the case if so much cold dust is present), and as such is not probed by the X-ray emission, then those authors' assumed gas masses are too low. Secondly, as the IR SED is fit by varying the timescale of shock evolution, assuming propagation through a uniform medium, the values are skewed towards those which reproduce the far-IR emission, which generally coincide with less dust destruction. The dust masses are thus higher than the actual values for the X-ray emitting gas, which, combined with a gas mass accounting only for this material, leads to an overestimate of the DTG ratio and the inferred surviving fraction of dust.

These parameters do not necessarily correspond to those of the actual shocked gas - \citet{arendt2010} find that the gas temperature inferred from the IR emission is lower than that directly measured from X-ray model fits, and \citet{temim2012b} also require a colder gas component to reproduce the IR flux, although they associate this with a cooler X-ray emitting component. This also explains the systematically higher destruction efficiencies found by measuring pre- and post-shock abundances from X-ray/UV/optical emission lines, which would predominantly be produced in the hot gas where significant dust destruction has occurred. The detection of high-velocity undepleted gas from absorption line studies (e.g. \citealt{jenkins1976}) is more problematic, as these measurements should be largely unaffected by the gas properties, but a small enough filling factor for the cold component, with a correspondingly low probability of intersecting the line of sight to a background star, would alleviate this tension.

{ The absorption measurements are also in much better agreement with the almost complete destruction of grains in the hot gas which we infer, compared to the only partial destruction found by previous IR studies. Theoretical models generally predict that $> 10\%$ of the initial dust mass survives, regardless of the details of the modelling or input parameters. While our estimate for the warm dust mass of \gtwentynine{} suggests a surviving fraction of $\sim 50 \%$, consistent with model predictions, the density of the hot gas in this object is noticeably lower than \geleven{} and \gtwentyseven{}, and suggests a preshock density closer to those assumed by e.g. \citet{slavin2015}. It is generally expected that higher gas densities result in higher destruction efficiencies, which is entirely consistent with the low surviving dust masses in the X-ray emitting regions of the other two SNRs, but at some point this trend must reverse as the shock cannot heat the ambient gas to the temperatures required for efficient sputtering of grains. This appears to be the case for the majority of the gas in all three SNRs, in line with expectations from models of substructure in molecular clouds \citep[e.g.][]{elmegreen2002}.}

\subsection{Implications}

While previously determined values of the dust destruction efficiency appear to be essentially correct, the interpretation and implied consequences are very different. If the dust is located in the hot gas, even if at present $50\%$ of the original mass has survived, destruction will continue until the gas has cooled too much for sputtering to be effective. The implied final destruction efficiency is therefore far higher, leading to the assumption that SNe can effectively clear the gas they sweep up of dust \citep{mckee1987}. If the dust is instead located in cold, dense gas where sputtering is ineffective, the eventual surviving fraction depends on whether this gas can be dispersed and the dust injected into hotter surroundings. If dispersal is inefficient, the final dust destruction efficiency may be close to the current value in the dense gas, the effective gas mass `cleared' of dust could be much lower than typically assumed, and the dust destruction timescale correspondingly longer. The tendency for large-scale simulations of the ISM to require highly efficient grain growth (e.g. \citealt{zhukovska2018}) to match observations may be a sign that their destruction timescales are in fact too low. A recent study by \citet{delooze2020} found that observed galaxy scaling relations are better fit by longer grain lifetimes and less efficient grain growth, supporting this interpretation. { \citet{gall2018} have also suggested that the relationship between dust and stellar masses can be explained by inefficient dust destruction processes, although they offer rapid replenishment of dust in the ISM as an alternative interpretation.}

This overestimation of the destruction efficiency would only be the case if the objects we investigate are representative of SNRs in general, which is not necessarily true given that we were able to select only three from a sample size of hundreds. However, while the number of SNRs showing clear shell structures in both the IR and X-ray is limited, \citet{chawner2019,chawner2020} find a much larger number with spatially correlated, if less regular, emission, which is suggestive of similar interactions with dense gas as the three studied in this paper. Broad CO emission is frequently detected towards SNRs \citep{kilpatrick2016}, suggesting that dense environments are far from unusual. \citet{koo2016} and \citet{chawner2020} have both studied large samples of SNRs and found that the IR SEDs are often poorly fit by homogeneous shock models, and a recent study of the Tornado SNR \citep{chawner2020b} finds the far-IR SED inconsistent with collisional heating of dust by the X-ray emitting gas, much as we have found here. It is at least plausible that a significant fraction of SNe explode in environments where their dust destruction efficiency is much lower than would be the case in a homogeneous ISM.

\subsection{Caveats}

{ While we interpret our results as indicating large masses of cold, swept-up gas and dust in all three SNRs, there are other possible explanations which do not have the same consequences for the dust destruction efficiency. As we show in Appendix \ref{sec:background}, despite the uncertainties associated with background subtraction, the far-IR excess is persistent and irreconcileable with our calculated collisionally-heated dust temperatures. Even ignoring the data longwards of $100 \um$ where ISM contamination becomes significant, the $70 \um$ flux is too high in two out of three objects, and is clearly morphologically associated with the SNR rather than background emission. We thus consider the need for an additional colder dust component to be firmly established. Based on the peak of the far-IR SEDs, the temperature of this cold dust cannot be much warmer than $30 \kel$, which, for typical dust opacities at these wavelengths, means the mass must be $> 0.1 \msun$ in order to reproduce the observed fluxes. This can be inferred directly from the observations with no modelling assumptions required, so is again a firm conclusion.}

{ We associate this additional cold dust with an undetected reservoir of cold gas which has passed through the shockwave. As mentioned above, \citet{temim2012b} instead assume the excess far-IR emission comes from a second X-ray emitting gas component with lower density and temperature in \gtwentynine{}, and we cannot rule out a similar interpretation for \geleven{} and \gtwentyseven{}. However, we consider this possibility unlikely for a number of reasons. Firstly, the densities we derive listed in Table \ref{tab:gasprop} are effectively lower limits, as they assume a filling factor of unity for the X-ray emitting gas. If the actual factor is lower, the density must be higher to compensate, which will raise the dust temperature and worsen the discrepency with the far-IR SED. Secondly, if the cold dust is located in cooler plasma, the mass of this gas component must be $> 10 \msun$ to remain consistent with the ISM DTG ratio. This is larger than the masses derived in Table \ref{tab:gasprop}, so this material should dominate, or at least significantly contribute, to the X-ray spectrum. There is no evidence for such material in the spectra of either \geleven{} or \gtwentyseven{}, and only tentative evidence in \gtwentynine{} (we are able to adequately fit the spectrum without including it). Finally, the temperature of shocked gas increases with decreasing preshock density, which makes the presence of gas which is both cooler {\it and} less dense than the bulk of the X-ray emitting material, as required by \citet{temim2012b}, difficult to explain.}

{ If we accept that the cold dust emission is coming from cold gas, it does not necessarily follow that this is swept-up material; fast shocks generate strong UV radiation which is capable of heating the gas and dust upstream \citep{docenko2010}, so we may be detecting material which is yet to be shock processed, and would presumably have a `surviving' dust fraction of nearly unity (small grains may be sublimated if heated to high enough temperatures). It is difficult to fully discount this possibility, but we regard it as less plausible - the emission from the blastwave in Cas A heats dust grains to $\sim 30 \kel$ at a distance of $1 \pc$ \citep{priestley2019}, so the dust within our apertures, which is presumably much closer than this to the heating source, should be at even higher temperatures, and so cannot explain the far-IR SED. Additionally, the $24$ and $70 \um$ images clearly show that the emission is coincident with the shocked gas, rather than extending further than the the X-ray emission as would be expected from dust heated by radiation ahead of the blastwave. It is also unclear whether enough dust can be heated in this manner to explain the far-IR fluxes, as the preheated region in \citet{docenko2010} makes up a tiny fraction of the width of the shock.}

{ Finally, we note that shock models do, in fact, predict the existence of colder material as the shocked gas cools to $\sim 10^4 \kel$, which removes the need for the dust to be located in localised overdensities. The models presented in \citet{slavin2015} suggest a density of $\sim 1 \pcc$ for this material, far lower than that inferred from (and necessary for) H$_2$ emission in \geleven{}. Moreover, this is inconsistent with the very high destruction efficiencies we find for the hot gas, unless a large fraction of the dust reforms in the post-shock material. This has been suggested to occur in SN1987A \citep{matsuura2019}, but would represent an even more extreme challenge to current models than an inhomogeneous ISM does, and likely result in an even greater increase in grain lifetimes. As with dust located ahead of the shock, it is also unclear whether enough of this material can exist to explain the large dust masses required by the far-IR SEDs. We thus regard high-density shocked gas as the most plausible interpretation of our results.}

\section{Conclusions}

We have demonstrated, for three SNRs, that the observed IR SEDs from the swept-up ISM are inconsistent with dust grains collisionally heated by the X-ray emitting gas, as the grain temperatures under these conditions are too high. The far-IR emission must be produced by a cold dust component with a mass of $\gtrsim 0.1 \msun$, compared to $\sim 10^{-3} \msun$ of dust in the hot gas. This cold dust must be associated with a mass of cold gas similarly larger than that of the X-ray emitting material. Whereas the dust destruction efficiency in the X-ray emitting gas is $\gtrsim 90 \%$, a much higher fraction of the cold dust has survived the SN shockwave. For \geleven{}, based on cold gas properties derived from H$_2$ emission lines, the minimum surviving fraction of dust { assuming a very conservative filling factor of unity} is $20 \%$ for carbon grains and $80 \%$ for silicates. As the total swept-up dust mass is dominated by the cold dust, which { we argue must come} from a denser phase of gas not captured by one-dimensional models, dust destruction timescales derived from these models - and commonly used in larger-scale simulations - may be significantly overestimating the efficiency of supernovae at destroying dust.

\section*{Acknowledgements}

We are grateful to Tea Temim for advice on modelling the X-ray spectra. FDP is funded by the Science and Technology Facilities Council. HC and HLG acknowledges support from the European Research Council (ERC) grant COSMICDUST ERC-2014-CoG-647939. IDL acknowledges support from ERC starting grant 851622 DustOrigin. MJB acknowledges support from the ERC grant SNDUST ERC-2015-AdG-694520.

\section*{Data Availability}

The data underlying this article are publically available at the respective telescope archives: \url{cda.harvard.edu/chaser/} ({\it Chandra}); \url{sha.ipac.caltech.edu/applications/Spitzer/SHA/} ({\it Spitzer}); \url{archives.esac.esa.int/hsa/whsa/} ({\it Herschel}). The derived properties used in the modelling are available in the article. All software used is publicly available.




\bibliographystyle{mnras}
\bibliography{destruction}

\begin{thebibliography}{}
\makeatletter
\relax
\def\mn@urlcharsother{\let\do\@makeother \do\$\do\&\do\#\do\^\do\_\do\%\do\~}
\def\mn@doi{\begingroup\mn@urlcharsother \@ifnextchar [ {\mn@doi@}
  {\mn@doi@[]}}
\def\mn@doi@[#1]#2{\def\@tempa{#1}\ifx\@tempa\@empty \href
  {http://dx.doi.org/#2} {doi:#2}\else \href {http://dx.doi.org/#2} {#1}\fi
  \endgroup}
\def\mn@eprint#1#2{\mn@eprint@#1:#2::\@nil}
\def\mn@eprint@arXiv#1{\href {http://arxiv.org/abs/#1} {{\tt arXiv:#1}}}
\def\mn@eprint@dblp#1{\href {http://dblp.uni-trier.de/rec/bibtex/#1.xml}
  {dblp:#1}}
\def\mn@eprint@#1:#2:#3:#4\@nil{\def\@tempa {#1}\def\@tempb {#2}\def\@tempc
  {#3}\ifx \@tempc \@empty \let \@tempc \@tempb \let \@tempb \@tempa \fi \ifx
  \@tempb \@empty \def\@tempb {arXiv}\fi \@ifundefined
  {mn@eprint@\@tempb}{\@tempb:\@tempc}{\expandafter \expandafter \csname
  mn@eprint@\@tempb\endcsname \expandafter{\@tempc}}}

\bibitem[\protect\citeauthoryear{{Andersen}, {Rho}, {Reach}, {Hewitt}  \&
  {Bernard}}{{Andersen} et~al.}{2011}]{andersen2011}
{Andersen} M.,  {Rho} J.,  {Reach} W.~T.,  {Hewitt} J.~W.,   {Bernard} J.~P.,
  2011, \mn@doi [\apj] {10.1088/0004-637X/742/1/7}, \href
  {https://ui.adsabs.harvard.edu/abs/2011ApJ...742....7A} {742, 7}

\bibitem[\protect\citeauthoryear{{Arendt} et~al.,}{{Arendt}
  et~al.}{2010}]{arendt2010}
{Arendt} R.~G.,  et~al., 2010, \mn@doi [\apj] {10.1088/0004-637X/725/1/585},
  \href {https://ui.adsabs.harvard.edu/abs/2010ApJ...725..585A} {725, 585}

\bibitem[\protect\citeauthoryear{{Balog} et~al.,}{{Balog}
  et~al.}{2014}]{balog2014}
{Balog} Z.,  et~al., 2014, \mn@doi [Experimental Astronomy]
  {10.1007/s10686-013-9352-3}, \href
  {https://ui.adsabs.harvard.edu/abs/2014ExA....37..129B} {37, 129}

\bibitem[\protect\citeauthoryear{{Barlow} \& {Silk}}{{Barlow} \&
  {Silk}}{1977}]{barlow1977}
{Barlow} M.~J.,  {Silk} J.,  1977, \mn@doi [\apjl] {10.1086/182346}, \href
  {https://ui.adsabs.harvard.edu/abs/1977ApJ...211L..83B} {211, L83}

\bibitem[\protect\citeauthoryear{{Bendo} et~al.,}{{Bendo}
  et~al.}{2013}]{bendo2013}
{Bendo} G.~J.,  et~al., 2013, \mn@doi [\mnras] {10.1093/mnras/stt948}, \href
  {https://ui.adsabs.harvard.edu/abs/2013MNRAS.433.3062B} {433, 3062}

\bibitem[\protect\citeauthoryear{{Bertoldi}, {Carilli}, {Cox}, {Fan},
  {Strauss}, {Beelen}, {Omont}  \& {Zylka}}{{Bertoldi}
  et~al.}{2003}]{bertoldi2003}
{Bertoldi} F.,  {Carilli} C.~L.,  {Cox} P.,  {Fan} X.,  {Strauss} M.~A.,
  {Beelen} A.,  {Omont} A.,   {Zylka} R.,  2003, \mn@doi [\aap]
  {10.1051/0004-6361:20030710}, \href
  {http://adsabs.harvard.edu/abs/2003A%26A...406L..55B} {406, L55}

\bibitem[\protect\citeauthoryear{{Bevan} \& {Barlow}}{{Bevan} \&
  {Barlow}}{2016}]{bevan2016}
{Bevan} A.,  {Barlow} M.~J.,  2016, \mn@doi [\mnras] {10.1093/mnras/stv2651},
  \href {http://adsabs.harvard.edu/abs/2016MNRAS.456.1269B} {456, 1269}

\bibitem[\protect\citeauthoryear{{Bevan} et~al.,}{{Bevan}
  et~al.}{2019}]{bevan2019}
{Bevan} A.,  et~al., 2019, \mn@doi [\mnras] {10.1093/mnras/stz679}, \href
  {https://ui.adsabs.harvard.edu/abs/2019MNRAS.485.5192B} {485, 5192}

\bibitem[\protect\citeauthoryear{{Borkowski} et~al.,}{{Borkowski}
  et~al.}{2006}]{borkowski2006}
{Borkowski} K.~J.,  et~al., 2006, \mn@doi [\apjl] {10.1086/504472}, \href
  {https://ui.adsabs.harvard.edu/abs/2006ApJ...642L.141B} {642, L141}

\bibitem[\protect\citeauthoryear{{Borkowski}, {Reynolds}  \&
  {Roberts}}{{Borkowski} et~al.}{2016}]{borkowski2016}
{Borkowski} K.~J.,  {Reynolds} S.~P.,   {Roberts} M.~S.~E.,  2016, \mn@doi
  [\apj] {10.3847/0004-637X/819/2/160}, \href
  {http://adsabs.harvard.edu/abs/2016ApJ...819..160B} {819, 160}

\bibitem[\protect\citeauthoryear{{Chawner} et~al.,}{{Chawner}
  et~al.}{2019}]{chawner2019}
{Chawner} H.,  et~al., 2019, \mn@doi [\mnras] {10.1093/mnras/sty2942}, \href
  {http://adsabs.harvard.edu/abs/2019MNRAS.483...70C} {483, 70}

\bibitem[\protect\citeauthoryear{{Chawner} et~al.,}{{Chawner}
  et~al.}{2020a}]{chawner2020b}
{Chawner} H.,  et~al., 2020a, \mn@doi [\mnras] {10.1093/mnras/staa2925}, \href
  {https://ui.adsabs.harvard.edu/abs/2020MNRAS.tmp.2747C} {}

\bibitem[\protect\citeauthoryear{{Chawner} et~al.,}{{Chawner}
  et~al.}{2020b}]{chawner2020}
{Chawner} H.,  et~al., 2020b, \mn@doi [\mnras] {10.1093/mnras/staa221}, \href
  {https://ui.adsabs.harvard.edu/abs/2020MNRAS.493.2706C} {493, 2706}

\bibitem[\protect\citeauthoryear{{De Looze}, {Barlow}, {Swinyard}, {Rho},
  {Gomez}, {Matsuura}  \& {Wesson}}{{De Looze} et~al.}{2017}]{delooze2017}
{De Looze} I.,  {Barlow} M.~J.,  {Swinyard} B.~M.,  {Rho} J.,  {Gomez} H.~L.,
  {Matsuura} M.,   {Wesson} R.,  2017, \mn@doi [\mnras]
  {10.1093/mnras/stw2837}, \href
  {http://adsabs.harvard.edu/abs/2017MNRAS.465.3309D} {465, 3309}

\bibitem[\protect\citeauthoryear{{De Looze} et~al.,}{{De Looze}
  et~al.}{2019}]{delooze2019}
{De Looze} I.,  et~al., 2019, \mn@doi [\mnras] {10.1093/mnras/stz1533}, \href
  {https://ui.adsabs.harvard.edu/abs/2019MNRAS.488..164D} {488, 164}

\bibitem[\protect\citeauthoryear{{De Looze} et~al.,}{{De Looze}
  et~al.}{2020}]{delooze2020}
{De Looze} I.,  et~al., 2020, \mn@doi [\mnras] {10.1093/mnras/staa1496}, \href
  {https://ui.adsabs.harvard.edu/abs/2020MNRAS.496.3668D} {496, 3668}

\bibitem[\protect\citeauthoryear{{Docenko} \& {Sunyaev}}{{Docenko} \&
  {Sunyaev}}{2010}]{docenko2010}
{Docenko} D.,  {Sunyaev} R.~A.,  2010, \mn@doi [\aap]
  {10.1051/0004-6361/200810366}, \href
  {http://adsabs.harvard.edu/abs/2010A%26A...509A..59D} {509, A59}

\bibitem[\protect\citeauthoryear{{Dopita}, {Seitenzahl}, {Sutherland }, {Vogt},
  {Winkler}  \& {Blair}}{{Dopita} et~al.}{2016}]{dopita2016}
{Dopita} M.~A.,  {Seitenzahl} I.~R.,  {Sutherland } R.~S.,  {Vogt} F. P.~A.,
  {Winkler} P.~F.,   {Blair} W.~P.,  2016, \mn@doi [\apj]
  {10.3847/0004-637X/826/2/150}, \href
  {https://ui.adsabs.harvard.edu/abs/2016ApJ...826..150D} {826, 150}

\bibitem[\protect\citeauthoryear{{Dorschner}, {Begemann}, {Henning}, {Jaeger}
  \& {Mutschke}}{{Dorschner} et~al.}{1995}]{dorschner1995}
{Dorschner} J.,  {Begemann} B.,  {Henning} T.,  {Jaeger} C.,   {Mutschke} H.,
  1995, \aap, \href {http://adsabs.harvard.edu/abs/1995A%26A...300..503D} {300,
  503}

\bibitem[\protect\citeauthoryear{{Draine}}{{Draine}}{2011}]{draine2011}
{Draine} B.~T.,  2011, {Physics of the Interstellar and Intergalactic Medium}

\bibitem[\protect\citeauthoryear{{Dwek}}{{Dwek}}{1987}]{dwek1987}
{Dwek} E.,  1987, \mn@doi [\apj] {10.1086/165774}, \href
  {http://adsabs.harvard.edu/abs/1987ApJ...322..812D} {322, 812}

\bibitem[\protect\citeauthoryear{{Elmegreen}}{{Elmegreen}}{2002}]{elmegreen2002}
{Elmegreen} B.~G.,  2002, \mn@doi [\apj] {10.1086/324384}, \href
  {https://ui.adsabs.harvard.edu/abs/2002ApJ...564..773E} {564, 773}

\bibitem[\protect\citeauthoryear{{Engelbracht} et~al.,}{{Engelbracht}
  et~al.}{2007}]{engelbracht2007}
{Engelbracht} C.~W.,  et~al., 2007, \mn@doi [\pasp] {10.1086/521881}, \href
  {https://ui.adsabs.harvard.edu/abs/2007PASP..119..994E} {119, 994}

\bibitem[\protect\citeauthoryear{{Foreman-Mackey}, {Hogg}, {Lang}  \&
  {Goodman}}{{Foreman-Mackey} et~al.}{2013}]{foreman2013}
{Foreman-Mackey} D.,  {Hogg} D.~W.,  {Lang} D.,   {Goodman} J.,  2013, \mn@doi
  [\pasp] {10.1086/670067}, \href
  {http://adsabs.harvard.edu/abs/2013PASP..125..306F} {125, 306}

\bibitem[\protect\citeauthoryear{{Freeman}, {Doe}  \&
  {Siemiginowska}}{{Freeman} et~al.}{2001}]{freeman2001}
{Freeman} P.,  {Doe} S.,   {Siemiginowska} A.,  2001, {Sherpa: a
  mission-independent data analysis application}.
pp 76--87, \mn@doi{10.1117/12.447161}

\bibitem[\protect\citeauthoryear{{Fruscione} et~al.,}{{Fruscione}
  et~al.}{2006}]{fruscione2006}
{Fruscione} A.,  et~al., 2006, {CIAO: Chandra's data analysis system}.
p. 62701V, \mn@doi{10.1117/12.671760}

\bibitem[\protect\citeauthoryear{{Gall} \& {Hjorth}}{{Gall} \&
  {Hjorth}}{2018}]{gall2018}
{Gall} C.,  {Hjorth} J.,  2018, \mn@doi [\apj] {10.3847/1538-4357/aae520},
  \href {http://adsabs.harvard.edu/abs/2018ApJ...868...62G} {868, 62}

\bibitem[\protect\citeauthoryear{{Gall}, {Hjorth}  \& {Andersen}}{{Gall}
  et~al.}{2011}]{gall2011}
{Gall} C.,  {Hjorth} J.,   {Andersen} A.~C.,  2011, \mn@doi [\aapr]
  {10.1007/s00159-011-0043-7}, \href
  {http://adsabs.harvard.edu/abs/2011A%26ARv..19...43G} {19, 43}

\bibitem[\protect\citeauthoryear{{Gomez} et~al.,}{{Gomez}
  et~al.}{2012}]{gomez2012}
{Gomez} H.~L.,  et~al., 2012, \mn@doi [\apj] {10.1088/0004-637X/760/1/96},
  \href {http://adsabs.harvard.edu/abs/2012ApJ...760...96G} {760, 96}

\bibitem[\protect\citeauthoryear{{Green}}{{Green}}{2004}]{green2004}
{Green} D.~A.,  2004, Bulletin of the Astronomical Society of India, \href
  {http://adsabs.harvard.edu/abs/2004BASI...32..335G} {32, 335}

\bibitem[\protect\citeauthoryear{{Hu}, {Zhukovska}, {Somerville}  \&
  {Naab}}{{Hu} et~al.}{2019}]{hu2019}
{Hu} C.-Y.,  {Zhukovska} S.,  {Somerville} R.~S.,   {Naab} T.,  2019, \mn@doi
  [\mnras] {10.1093/mnras/stz1481}, \href
  {https://ui.adsabs.harvard.edu/abs/2019MNRAS.487.3252H} {487, 3252}

\bibitem[\protect\citeauthoryear{{Jarrett} et~al.,}{{Jarrett}
  et~al.}{2013}]{jarrett2013}
{Jarrett} T.~H.,  et~al., 2013, \mn@doi [\aj] {10.1088/0004-6256/145/1/6},
  \href {https://ui.adsabs.harvard.edu/abs/2013AJ....145....6J} {145, 6}

\bibitem[\protect\citeauthoryear{{Jenkins}, {Silk}  \& {Wallerstein}}{{Jenkins}
  et~al.}{1976}]{jenkins1976}
{Jenkins} E.~B.,  {Silk} J.,   {Wallerstein} G.,  1976, \mn@doi [\apjs]
  {10.1086/190412}, \href
  {https://ui.adsabs.harvard.edu/abs/1976ApJS...32..681J} {32, 681}

\bibitem[\protect\citeauthoryear{{Jones}, {Tielens}  \& {Hollenbach}}{{Jones}
  et~al.}{1996}]{jones1996}
{Jones} A.~P.,  {Tielens} A.~G.~G.~M.,   {Hollenbach} D.~J.,  1996, \mn@doi
  [\apj] {10.1086/177823}, \href
  {http://adsabs.harvard.edu/abs/1996ApJ...469..740J} {469, 740}

\bibitem[\protect\citeauthoryear{{Kilpatrick}, {Bieging}  \&
  {Rieke}}{{Kilpatrick} et~al.}{2016}]{kilpatrick2016}
{Kilpatrick} C.~D.,  {Bieging} J.~H.,   {Rieke} G.~H.,  2016, \mn@doi [\apj]
  {10.3847/0004-637X/816/1/1}, \href
  {https://ui.adsabs.harvard.edu/abs/2016ApJ...816....1K} {816, 1}

\bibitem[\protect\citeauthoryear{{Kirchschlager}, {Schmidt}, {Barlow},
  {Fogerty}, {Bevan}  \& {Priestley}}{{Kirchschlager}
  et~al.}{2019}]{kirchschlager2019}
{Kirchschlager} F.,  {Schmidt} F.~D.,  {Barlow} M.~J.,  {Fogerty} E.~L.,
  {Bevan} A.,   {Priestley} F.~D.,  2019, \mn@doi [\mnras]
  {10.1093/mnras/stz2399}, \href
  {https://ui.adsabs.harvard.edu/abs/2019MNRAS.489.4465K} {489, 4465}

\bibitem[\protect\citeauthoryear{{K{\"o}hler}, {Ysard}  \&
  {Jones}}{{K{\"o}hler} et~al.}{2015}]{kohler2015}
{K{\"o}hler} M.,  {Ysard} N.,   {Jones} A.~P.,  2015, \mn@doi [\aap]
  {10.1051/0004-6361/201525646}, \href
  {https://ui.adsabs.harvard.edu/abs/2015A&A...579A..15K} {579, A15}

\bibitem[\protect\citeauthoryear{{Koo}, {Moon}, {Lee}, {Lee}  \&
  {Matthews}}{{Koo} et~al.}{2007}]{koo2007}
{Koo} B.-C.,  {Moon} D.-S.,  {Lee} H.-G.,  {Lee} J.-J.,   {Matthews} K.,  2007,
  \mn@doi [\apj] {10.1086/510550}, \href
  {https://ui.adsabs.harvard.edu/abs/2007ApJ...657..308K} {657, 308}

\bibitem[\protect\citeauthoryear{{Koo}, {Lee}, {Jeong}, {Seok}  \& {Kim}}{{Koo}
  et~al.}{2016}]{koo2016}
{Koo} B.-C.,  {Lee} J.-J.,  {Jeong} I.-G.,  {Seok} J.~Y.,   {Kim} H.-J.,  2016,
  \mn@doi [\apj] {10.3847/0004-637X/821/1/20}, \href
  {https://ui.adsabs.harvard.edu/abs/2016ApJ...821...20K} {821, 20}

\bibitem[\protect\citeauthoryear{{Kumar}, {Safi-Harb}, {Slane}  \&
  {Gotthelf}}{{Kumar} et~al.}{2014}]{kumar2014}
{Kumar} H.~S.,  {Safi-Harb} S.,  {Slane} P.~O.,   {Gotthelf} E.~V.,  2014,
  \mn@doi [\apj] {10.1088/0004-637X/781/1/41}, \href
  {https://ui.adsabs.harvard.edu/abs/2014ApJ...781...41K} {781, 41}

\bibitem[\protect\citeauthoryear{{Mart{\'\i}nez-Gonz{\'a}lez}, {W{\"u}nsch},
  {Silich}, {Tenorio-Tagle}, {Palou{\v{s}}}  \&
  {Ferrara}}{{Mart{\'\i}nez-Gonz{\'a}lez} et~al.}{2019}]{martinezgonzales2019}
{Mart{\'\i}nez-Gonz{\'a}lez} S.,  {W{\"u}nsch} R.,  {Silich} S.,
  {Tenorio-Tagle} G.,  {Palou{\v{s}}} J.,   {Ferrara} A.,  2019, \mn@doi [\apj]
  {10.3847/1538-4357/ab571b}, \href
  {https://ui.adsabs.harvard.edu/abs/2019ApJ...887..198M} {887, 198}

\bibitem[\protect\citeauthoryear{{Mathis}, {Rumpl}  \& {Nordsieck}}{{Mathis}
  et~al.}{1977}]{mathis1977}
{Mathis} J.~S.,  {Rumpl} W.,   {Nordsieck} K.~H.,  1977, \mn@doi [\apj]
  {10.1086/155591}, \href {http://adsabs.harvard.edu/abs/1977ApJ...217..425M}
  {217, 425}

\bibitem[\protect\citeauthoryear{{Mathis}, {Mezger}  \& {Panagia}}{{Mathis}
  et~al.}{1983}]{mathis1983}
{Mathis} J.~S.,  {Mezger} P.~G.,   {Panagia} N.,  1983, \aap, \href
  {http://adsabs.harvard.edu/abs/1983A%26A...128..212M} {128, 212}

\bibitem[\protect\citeauthoryear{{Matsuura} et~al.,}{{Matsuura}
  et~al.}{2011}]{matsuura2011}
{Matsuura} M.,  et~al., 2011, \mn@doi [Science] {10.1126/science.1205983},
  \href {http://adsabs.harvard.edu/abs/2011Sci...333.1258M} {333, 1258}

\bibitem[\protect\citeauthoryear{{Matsuura} et~al.,}{{Matsuura}
  et~al.}{2019}]{matsuura2019}
{Matsuura} M.,  et~al., 2019, \mn@doi [\mnras] {10.1093/mnras/sty2734}, \href
  {https://ui.adsabs.harvard.edu/abs/2019MNRAS.482.1715M} {482, 1715}

\bibitem[\protect\citeauthoryear{{McKee}, {Hollenbach}, {Seab}  \&
  {Tielens}}{{McKee} et~al.}{1987}]{mckee1987}
{McKee} C.~F.,  {Hollenbach} D.~J.,  {Seab} G.~C.,   {Tielens} A.~G.~G.~M.,
  1987, \mn@doi [\apj] {10.1086/165403}, \href
  {https://ui.adsabs.harvard.edu/abs/1987ApJ...318..674M} {318, 674}

\bibitem[\protect\citeauthoryear{{Morgan} \& {Edmunds}}{{Morgan} \&
  {Edmunds}}{2003}]{morgan2003}
{Morgan} H.~L.,  {Edmunds} M.~G.,  2003, \mn@doi [\mnras]
  {10.1046/j.1365-8711.2003.06681.x}, \href
  {http://adsabs.harvard.edu/abs/2003MNRAS.343..427M} {343, 427}

\bibitem[\protect\citeauthoryear{{Nozawa}, {Kozasa}, {Habe}, {Dwek}, {Umeda},
  {Tominaga}, {Maeda}  \& {Nomoto}}{{Nozawa} et~al.}{2007}]{nozawa2007}
{Nozawa} T.,  {Kozasa} T.,  {Habe} A.,  {Dwek} E.,  {Umeda} H.,  {Tominaga} N.,
   {Maeda} K.,   {Nomoto} K.,  2007, \mn@doi [\apj] {10.1086/520621}, \href
  {http://adsabs.harvard.edu/abs/2007ApJ...666..955N} {666, 955}

\bibitem[\protect\citeauthoryear{{Priddey}, {Isaak}, {McMahon}, {Robson}  \&
  {Pearson}}{{Priddey} et~al.}{2003}]{priddey2003}
{Priddey} R.~S.,  {Isaak} K.~G.,  {McMahon} R.~G.,  {Robson} E.~I.,   {Pearson}
  C.~P.,  2003, \mn@doi [\mnras] {10.1046/j.1365-8711.2003.07076.x}, \href
  {http://adsabs.harvard.edu/abs/2003MNRAS.344L..74P} {344, L74}

\bibitem[\protect\citeauthoryear{{Priestley} \& {Barlow}}{{Priestley} \&
  {Barlow}}{2018}]{priestley2018}
{Priestley} F.~D.,  {Barlow} M.~J.,  2018, \mn@doi [\mnras]
  {10.1093/mnras/sty1099}, \href
  {https://ui.adsabs.harvard.edu/abs/2018MNRAS.478.1502P} {478, 1502}

\bibitem[\protect\citeauthoryear{{Priestley}, {Barlow}  \& {Viti}}{{Priestley}
  et~al.}{2017}]{priestley2017}
{Priestley} F.~D.,  {Barlow} M.~J.,   {Viti} S.,  2017, \mn@doi [\mnras]
  {10.1093/mnras/stx2327}, \href
  {http://adsabs.harvard.edu/abs/2017MNRAS.472.4444P} {472, 4444}

\bibitem[\protect\citeauthoryear{{Priestley}, {Barlow}  \& {De
  Looze}}{{Priestley} et~al.}{2019}]{priestley2019}
{Priestley} F.~D.,  {Barlow} M.~J.,   {De Looze} I.,  2019, \mn@doi [\mnras]
  {10.1093/mnras/stz414}, \href
  {http://adsabs.harvard.edu/abs/2019MNRAS.485..440P} {485, 440}

\bibitem[\protect\citeauthoryear{{Priestley}, {Barlow}, {De Looze}  \&
  {Chawner}}{{Priestley} et~al.}{2020}]{priestley2020}
{Priestley} F.~D.,  {Barlow} M.~J.,  {De Looze} I.,   {Chawner} H.,  2020,
  \mn@doi [\mnras] {10.1093/mnras/stz3434}, \href
  {https://ui.adsabs.harvard.edu/abs/2020MNRAS.491.6020P} {491, 6020}

\bibitem[\protect\citeauthoryear{{Ranasinghe} \& {Leahy}}{{Ranasinghe} \&
  {Leahy}}{2018}]{ranasinghe2018}
{Ranasinghe} S.,  {Leahy} D.~A.,  2018, \mn@doi [\aj]
  {10.3847/1538-3881/aab9be}, \href
  {https://ui.adsabs.harvard.edu/abs/2018AJ....155..204R} {155, 204}

\bibitem[\protect\citeauthoryear{{Reach}, {Rho}  \& {Jarrett}}{{Reach}
  et~al.}{2005}]{reach2005}
{Reach} W.~T.,  {Rho} J.,   {Jarrett} T.~H.,  2005, \mn@doi [\apj]
  {10.1086/425855}, \href
  {https://ui.adsabs.harvard.edu/abs/2005ApJ...618..297R} {618, 297}

\bibitem[\protect\citeauthoryear{{Sankrit} et~al.,}{{Sankrit}
  et~al.}{2010}]{sankrit2010}
{Sankrit} R.,  et~al., 2010, \mn@doi [\apj] {10.1088/0004-637X/712/2/1092},
  \href {https://ui.adsabs.harvard.edu/abs/2010ApJ...712.1092S} {712, 1092}

\bibitem[\protect\citeauthoryear{{Slavin}, {Dwek}  \& {Jones}}{{Slavin}
  et~al.}{2015}]{slavin2015}
{Slavin} J.~D.,  {Dwek} E.,   {Jones} A.~P.,  2015, \mn@doi [\apj]
  {10.1088/0004-637X/803/1/7}, \href
  {https://ui.adsabs.harvard.edu/abs/2015ApJ...803....7S} {803, 7}

\bibitem[\protect\citeauthoryear{{Temim}, {Slane}, {Arendt}  \& {Dwek}}{{Temim}
  et~al.}{2012}]{temim2012b}
{Temim} T.,  {Slane} P.,  {Arendt} R.~G.,   {Dwek} E.,  2012, \mn@doi [\apj]
  {10.1088/0004-637X/745/1/46}, \href
  {https://ui.adsabs.harvard.edu/abs/2012ApJ...745...46T} {745, 46}

\bibitem[\protect\citeauthoryear{{Verbiest}, {Weisberg}, {Chael}, {Lee}  \&
  {Lorimer}}{{Verbiest} et~al.}{2012}]{verbiest2012}
{Verbiest} J.~P.~W.,  {Weisberg} J.~M.,  {Chael} A.~A.,  {Lee} K.~J.,
  {Lorimer} D.~R.,  2012, \mn@doi [\apj] {10.1088/0004-637X/755/1/39}, \href
  {http://adsabs.harvard.edu/abs/2012ApJ...755...39V} {755, 39}

\bibitem[\protect\citeauthoryear{{Watson}, {Christensen}, {Knudsen}, {Richard},
  {Gallazzi}  \& {Micha{\l}owski}}{{Watson} et~al.}{2015}]{watson2015}
{Watson} D.,  {Christensen} L.,  {Knudsen} K.~K.,  {Richard} J.,  {Gallazzi}
  A.,   {Micha{\l}owski} M.~J.,  2015, \mn@doi [\nat] {10.1038/nature14164},
  \href {http://adsabs.harvard.edu/abs/2015Natur.519..327W} {519, 327}

\bibitem[\protect\citeauthoryear{{Wesson}, {Barlow}, {Matsuura}  \&
  {Ercolano}}{{Wesson} et~al.}{2015}]{wesson2015}
{Wesson} R.,  {Barlow} M.~J.,  {Matsuura} M.,   {Ercolano} B.,  2015, \mn@doi
  [\mnras] {10.1093/mnras/stu2250}, \href
  {http://adsabs.harvard.edu/abs/2015MNRAS.446.2089W} {446, 2089}

\bibitem[\protect\citeauthoryear{{Williams} et~al.,}{{Williams}
  et~al.}{2006}]{williams2006}
{Williams} B.~J.,  et~al., 2006, \mn@doi [\apjl] {10.1086/509876}, \href
  {https://ui.adsabs.harvard.edu/abs/2006ApJ...652L..33W} {652, L33}

\bibitem[\protect\citeauthoryear{{Ysard}, {K{\"o}hler}, {Jones}, {Dartois},
  {Godard}  \& {Gavilan}}{{Ysard} et~al.}{2016}]{ysard2016}
{Ysard} N.,  {K{\"o}hler} M.,  {Jones} A.,  {Dartois} E.,  {Godard} M.,
  {Gavilan} L.,  2016, \mn@doi [\aap] {10.1051/0004-6361/201527487}, \href
  {https://ui.adsabs.harvard.edu/abs/2016A&A...588A..44Y} {588, A44}

\bibitem[\protect\citeauthoryear{{Zhu}, {Tian}  \& {Zuo}}{{Zhu}
  et~al.}{2014}]{zhu2014}
{Zhu} H.,  {Tian} W.~W.,   {Zuo} P.,  2014, \mn@doi [\apj]
  {10.1088/0004-637X/793/2/95}, \href
  {https://ui.adsabs.harvard.edu/abs/2014ApJ...793...95Z} {793, 95}

\bibitem[\protect\citeauthoryear{{Zhu}, {Slane}, {Raymond}  \& {Tian}}{{Zhu}
  et~al.}{2019}]{zhu2019}
{Zhu} H.,  {Slane} P.,  {Raymond} J.,   {Tian} W.~W.,  2019, \mn@doi [\apj]
  {10.3847/1538-4357/ab3226}, \href
  {https://ui.adsabs.harvard.edu/abs/2019ApJ...882..135Z} {882, 135}

\bibitem[\protect\citeauthoryear{{Zhukovska}, {Henning}  \&
  {Dobbs}}{{Zhukovska} et~al.}{2018}]{zhukovska2018}
{Zhukovska} S.,  {Henning} T.,   {Dobbs} C.,  2018, \mn@doi [\apj]
  {10.3847/1538-4357/aab438}, \href
  {https://ui.adsabs.harvard.edu/abs/2018ApJ...857...94Z} {857, 94}

\bibitem[\protect\citeauthoryear{{Zubko}, {Mennella}, {Colangeli}  \&
  {Bussoletti}}{{Zubko} et~al.}{1996}]{zubko1996}
{Zubko} V.~G.,  {Mennella} V.,  {Colangeli} L.,   {Bussoletti} E.,  1996,
  \mn@doi [\mnras] {10.1093/mnras/282.4.1321}, \href
  {http://adsabs.harvard.edu/abs/1996MNRAS.282.1321Z} {282, 1321}

\makeatother
\end{thebibliography}


\appendix

\section{Background subtraction}
\label{sec:background}

\begin{figure}
  \centering
  \includegraphics[width=\columnwidth]{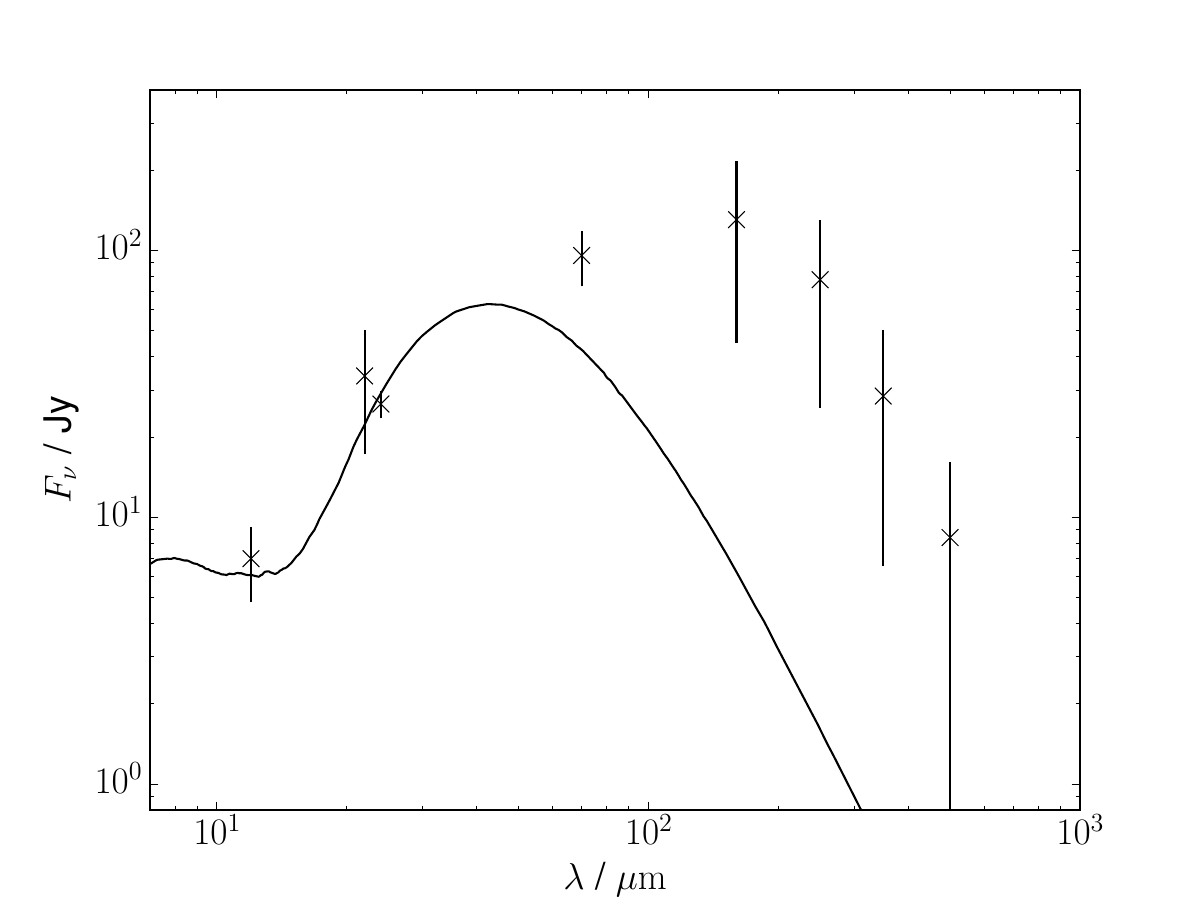}
  \caption{Best-fit collisionally heated dust SED for \geleven{} using carbon grains (solid line) and observed fluxes (crosses), using an alternative backround subtraction.}
  \label{fig:altg11}
\end{figure}

\begin{figure}
  \centering
  \includegraphics[width=\columnwidth]{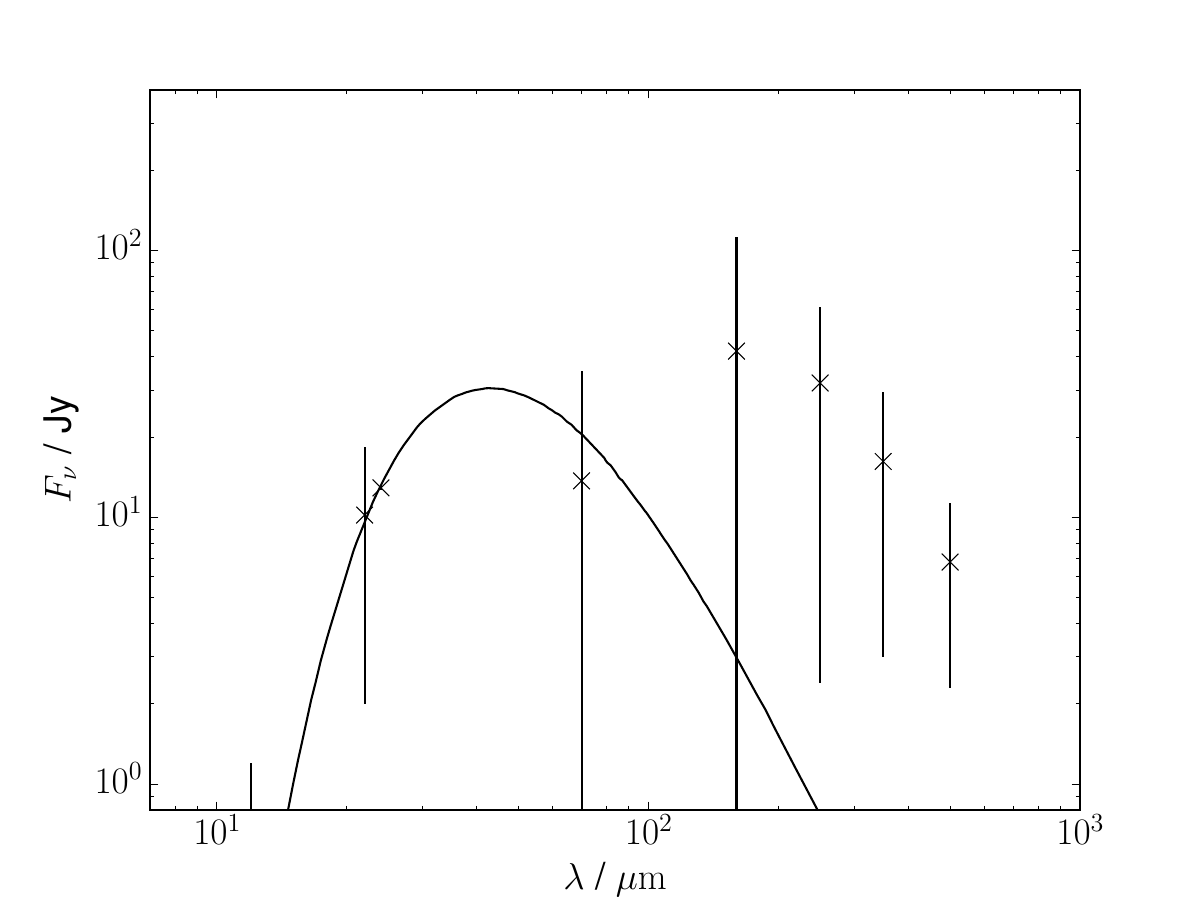}
  \caption{Best-fit collisionally heated dust SED for \gtwentyseven{} using carbon grains (solid line) and observed fluxes (crosses), using an alternative backround subtraction.}
  \label{fig:altg27}
\end{figure}

\begin{figure}
  \centering
  \includegraphics[width=\columnwidth]{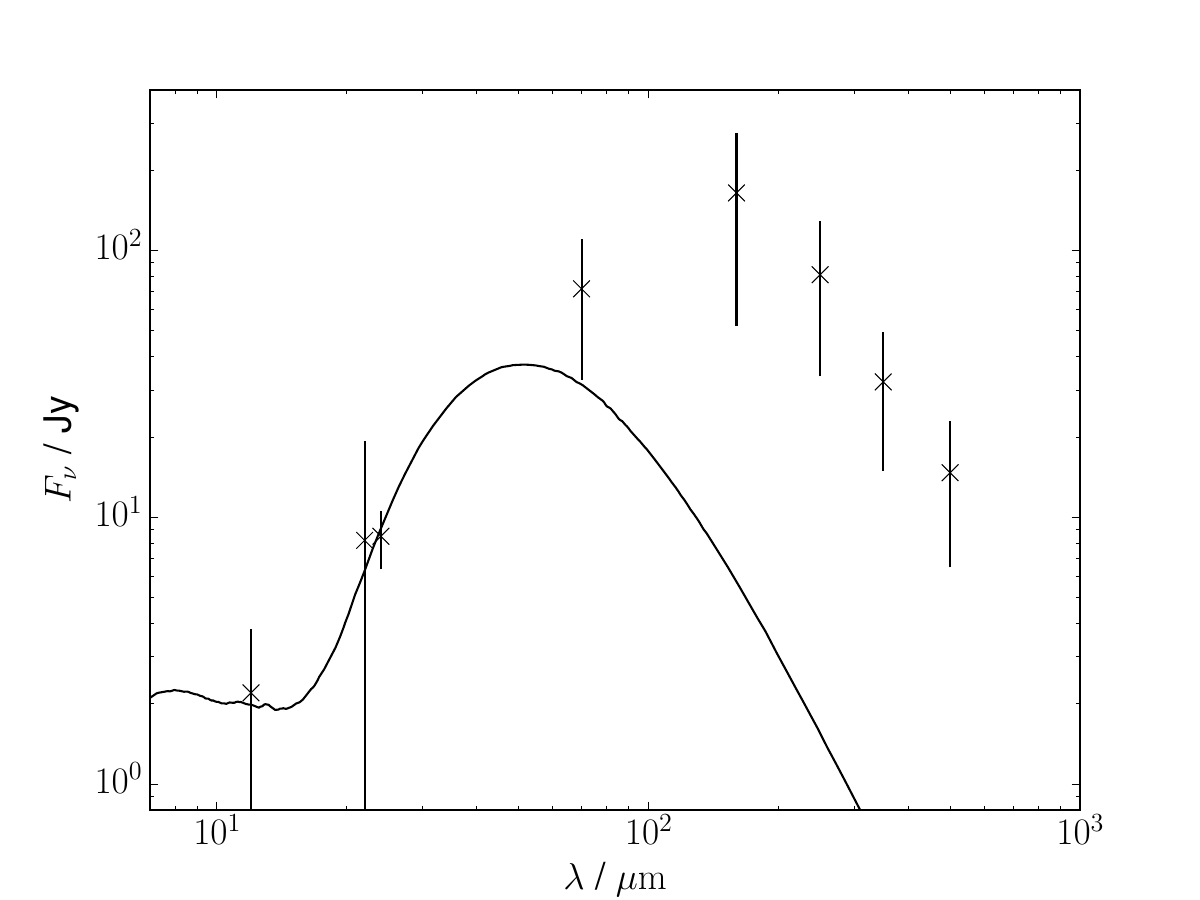}
  \caption{Best-fit collisionally heated dust SED for \gtwentynine{} using carbon grains (solid line) and observed fluxes (crosses), using an alternative backround subtraction.}
  \label{fig:altg29}
\end{figure}

In determining the fluxes listed in Table \ref{tab:irflux} we subtracted a background flux for each filter, defined as the median value from a nearby region selected to be `typical' of the local environment. As this is a largely subjective choice, it is possible that our assumed fluxes are biased, potentially in either direction and not necessarily in the same way for all filters. Table \ref{tab:altflux} lists fluxes where, rather than a particular region, we instead take the median of the entire image as the background contribution. Figures \ref{fig:altg11}, \ref{fig:altg27} and \ref{fig:altg29} show the resulting SED fits with carbon grains for \geleven{}, \gtwentyseven{} and \gtwentynine{} respectively. For \geleven{}, there is still a clear far-IR excess which cannot be explained by grains located in the hot gas. For \gtwentynine{}, while the $70 \um$ flux is now just about consistent with warm dust, there is a statistically significant excess at longer wavelengths which was not present using the previous background subtraction. For \gtwentyseven{}, the evidence for a far-IR excess is weaker, but still present. Clearly, the assumed background can have a significant impact on the far-IR fluxes, and thus the dust mass of any additional colder component. However, in all cases there is still a need for the presence of colder dust. Given the uncertainties in the heating mechanism, we are unable to reliably constrain the cold dust mass even if we had perfect knowledge of the background fluxes, but this does not affect our conclusion that the mass in this component is necessarily much larger than the warm dust heated by the hot gas.

\begin{table*}
  \centering
  \caption{Alternative background-subtracted IR fluxes extracted from apertures defined in Table \ref{tab:snrs}. All fluxes are in Jy, and band wavelengths are given in $\um$.}
  \begin{tabular}{ccccccccc}
    \hline
    SNR & WISE $12$ & WISE $22$ & MIPS $24$ & PACS $70$ & PACS $160$ & SPIRE $250$ & SPIRE $350$ & SPIRE $500$ \\
    \hline
    \geleven & $7.0 \pm 2.2$ & $33.9 \pm 16.6$ & $26.6 \pm 3.1$ & $95.9 \pm 22.4$ & $130.6 \pm 85.6$ & $77.8 \pm 52.1$ & $28.5 \pm 21.9$ & $8.4 \pm 7.7$ \\
    \gtwentyseven & $0.0 \pm 1.2$ & $10.2 \pm 8.2$ & $12.9 \pm 0.2$ & $13.8 \pm 21.6$ & $42.0 \pm 70.0$ & $31.9 \pm 29.5$ & $16.2 \pm 13.2$ & $6.8 \pm 4.5$ \\
    \gtwentynine & $2.2 \pm 1.6$ & $8.2 \pm 11.2$ & $8.5 \pm 2.1$ & $71.9 \pm 39.1$ & $164.5 \pm 112.4$ & $81.2 \pm 47.4$ & $32.2 \pm 17.3$ & $14.8 \pm 8.2$ \\
    \hline
  \end{tabular}
  \label{tab:altflux}
\end{table*}

\bsp	
\label{lastpage}
\end{document}